\documentclass[prb,twocolumn,amsmath,amssymb,nofootinbib]{revtex4}
\usepackage{graphicx}
\usepackage{amsmath}
\usepackage{amssymb}
\usepackage{color}
\usepackage{hyperref}
\usepackage{amsthm}
\usepackage{booktabs}

\def\bea{\begin{eqnarray}}
\def\eea{\end{eqnarray}}
\def\<{\langle}
\def\>{\rangle}
\def\tr{\text{Tr}}
\def\be{\begin{equation}}
\def\ee{\end{equation}}
\def\non{\nonumber}
\def\ep{\text{ep}}
\def\avg#1{\overline{#1}}
\def\Id{\mathbf{1}}
\def\CC{\mathbb{C}}
\def\dd{\text{d}}

\begin{document}

\setcitestyle{square,numbers,comma}
\makeatletter
\def\@biblabel#1{[#1]}
\makeatother

\title{Entangling Power: A Probe of Symmetry and Integrability \\ in Quantum Many-Body Systems}

\author{Ian Low\footnote{Corresponding author: \texttt{ilow@northwestern.edu}}, Pallab Goswami}
\affiliation{\vspace{0.1cm}
\mbox{Department of Physics and Astronomy, Northwestern University, Evanston, IL 60208, USA}\\
\vspace{-0.3cm}}

\begin{abstract}
The entangling power of a unitary operator quantifies its ability to generate entanglement from product states and provides a natural probe of quantum many-body dynamics. Entanglement extremization at points of enhanced symmetry has previously been observed in high-energy scattering.
In this work we compute the time-averaged entangling power of anisotropic Heisenberg spin chains across two-site models and finite-size systems, as well as the entangling power of the two-magnon
$S$-matrix in the thermodynamic limit. For two-site models we establish a monotonic hierarchy: the entangling power decreases as the symmetry group grows, reaching its minimum at the $SU(2)$ XXX point. Finite-size XXZ chains exhibit sharp dips at the $SU(2)$ points $\Delta=\pm 1$ and the free-fermion point $\Delta=0$, with the free-fermion dip decaying much more slowly with system size. In the thermodynamic limit, we decompose the two-magnon $S$-matrix into quantum logic gates---Identity, SWAP, and $\sigma_z\otimes\sigma_z$---and show that the entangling power vanishes for all scattering energies at the $SU(2)$ points, where the $S$-matrix reduces to the Identity gate, while the free-fermion point achieves the maximum--- the opposite of the finite-size many-body behavior. The entangling power can serve as an {\em operator} diagnostic for symmetry and selected aspects of integrability in quantum simulations of  spin-chain dynamics.
\end{abstract}

\maketitle

\section{Introduction}
\label{sec:intro}

Symmetry is arguably the most powerful organizing principle
in physics. From conservation laws and selection rules to the
classification of fundamental interactions and phases of matter,
symmetry considerations pervade nearly every branch of physics. 
Yet the origin of symmetry remains one of the deepest
open questions and its existence always postulated rather than derived. 
Understanding {\em why} certain
symmetries appear in nature, rather than merely categorizing
them, is a challenge that may require fundamentally new
organizing principles.

Recent developments at the intersection of quantum information science and fundamental physics provide a tantalizing hint that such a principle may exist. In low-energy nucleon-nucleon scattering, the entangling power of the
 $S$-matrix is minimized at the Wigner $SU(4)$ point, where spin and isospin symmetries combine into a larger group, as well as the unitarity limit when non-relativistic conformal invariance appears \cite{Beane:2018oxh,Low:2021ufv}. Similar correlations between entanglement extremization and enhanced symmetry have been identified in a growing number of scattering processes under different contexts, including low-energy hadron dynamics \cite{Liu:2022grf,Liu:2024frx,Hu:2025lua}, beyond-the-Standard-Model extensions for the Higgs boson \cite{Carena:2023vjc, Carena:2025wyh, Busoni:2025dns, Li:2026kha}, and the general S-matrix framework \cite{McGinnis:2025brt,McGinnis:2025xgt}. These findings suggest that symmetry may emerge from an extremization principle rooted in quantum entanglement, offering a new perspective on its origin. In addition, further investigations suggest that extremization of quantum resources may predict fundamental constants in nature, such as the Cabibbo-Kobayashi-Maskawa matrix \cite{Thaler:2024anb}, weak mixing angle \cite{Liu:2025bgw}, and the Higgs  mass \cite{Liu:2025iwh}.

In quantum many-body physics, the interplay of entanglement and
symmetry has been explored extensively, but through a
rather different lens. Ground-state entanglement
entropy~\cite{Calabrese:2004eu,Calabrese:2009qy}
and its area-law
scaling~\cite{Vidal:2002rm,Latorre:2003kg} have become central tools for
characterizing quantum phases and critical
points~\cite{Osterloh:2002na,Osborne:2002zz};
see Refs.~\cite{Amico:2007ag,Eisert:2008ur} for reviews.
Symmetry-resolved entanglement, which decomposes the
entanglement entropy into contributions from 
 sectors labeled by conserved charges, has attracted considerable attention
as a refined
probe~\cite{Goldstein:2017bua,Xavier:2018eua,Murciano:2020lqq,Ares:2022koq}.
These are powerful frameworks, but they share two
limitations: they characterize a particular {\em state}
(typically the ground state) rather than the dynamics
itself, and symmetry-resolved entanglement requires
knowing {\em a priori} the existence of conserved charges and, therefore, the symmetry.


In this work we take a fundamentally different approach.
The entangling power, introduced in Refs.~\cite{Zanardi:2000zz,Zanardi:2001qu}, is a property of the
{\em operator} $U$ rather than any single state---it
measures the average entanglement generated by $U$
acting on product states drawn uniformly from the Haar
measure.
Crucially, no assumption about the symmetry of the
system is needed; instead, the symmetry reveals itself
through the behavior of the entangling power.
When applied to the time-evolution operator
$U(t) = e^{iHt}$ of a many-body Hamiltonian, the
time-averaged entangling power $\avg{\ep}$ captures
the typical entanglement production of the
dynamics~\cite{Pal:2018epsc,Pal:2024xlp} and
provides a symmetry probe complementary to existing
state-level diagnostics.
We compute $\avg{\ep}$ for anisotropic Heisenberg spin
chains across three regimes that are usually studied
separately---two-site quantum gates, finite-size
many-body chains, and the two-magnon $S$-matrix in the
thermodynamic limit---and show that in every regime
the entangling power is suppressed at points of
enhanced algebraic structure.
The entangling power thus functions as a dynamical
symmetry witness across scales, with practical
implications for quantum simulation and Hamiltonian
characterization: it diagnoses the symmetry content
of the dynamics without reference to any particular
eigenstate or ground state.

This paper is organized as follows.
In Sec.~\ref{sec:ep} we review the entangling power
and its time-averaged version.
In Sec.~\ref{sec:twosite} we present the two-site
results, including the symmetry hierarchy, eigenvalue
analysis, and isospectral decomposition.
In Sec.~\ref{sec:finitesize} we extend to finite-size
spin-$1/2$ and spin-$1$ chains and study the persistence
and decay of the symmetry dip, the emergence and mechanism
of the free-fermion dip, and their contrasting scaling
with system size.
In Sec.~\ref{sec:integrability} we disentangle the roles
of symmetry and integrability by studying the $J_1$-$J_2$
model and the bilinear-biquadratic spin-$1$ chain.
In Sec.~\ref{sec:twomagnon} we introduce the six-vertex
$R$-matrix that encodes two-magnon scattering in the
XXZ chain, compute its entangling power as a function
of rapidity and anisotropy, decompose it into quantum
logic gates, and show that the correlation
between entanglement suppression and symmetry
enhancement extends to the thermodynamic limit.
We conclude with a discussion in
Sec.~\ref{sec:discussion}.
A pedagogical derivation of the Bethe ansatz and the
two-magnon $S$-matrix is provided in
Appendix~\ref{app:bethe}.

\section{Entangling Power}
\label{sec:ep}

Quantum entanglement is ordinarily a property of a
state: a given state vector in a bipartite Hilbert
space ${\cal H}_A \otimes {\cal H}_B$, its degree of entanglement is
quantified by an entanglement measure applied to the
state.
The entanglement {\em generated} by a unitary operator
$U$, however, depends on the input state it acts on.
A familiar example is the CNOT gate: acting on the
computational-basis state $|0\> \otimes |0\>$ it
produces the product state $|0\> \otimes |0\>$, while
acting on $|+\> \otimes |0\>$ (with
$|+\> = (|0\> + |1\>)/\sqrt{2}$) it produces the
maximally entangled Bell state
$(|00\> + |11\>)/\sqrt{2}$.
To characterize the entangling capability of  a unitary operator, one 
averages over input states.

The entangling power, introduced in
Refs.~\cite{Zanardi:2000zz,Zanardi:2001qu}, does
exactly this.
Given a bipartite unitary $U$ acting on
${\cal H}_A \otimes {\cal H}_B$ with
$d_i = \dim{\cal H}_i$, one draws random states
$|\psi\> \in {\cal H}_A$ and $|\phi\> \in {\cal H}_B$
independently from the Haar measure, applies $U$ to the
product state $|\psi\> \otimes |\phi\>$, and measures
the entanglement of the output via the reduced density
matrix
$\rho_A = \tr_B\bigl(
U|\psi,\phi\>\<\psi,\phi|U^\dagger\bigr)$.
One must choose an entanglement measure; following
Refs.~\cite{Zanardi:2000zz,Zanardi:2001qu} we adopt
the linear entropy~\cite{Nielsen:2012yss}
${\cal E}(|\psi\>) = 1 - \tr(\rho_A^2)$,
which vanishes when $|\psi\>$ is a product state
and is maximal for maximally mixed
$\rho_A$~\cite{Hu:2025lua}.
The entangling power is then defined as the Haar average
of the linear entropy of the output states over all
product-state inputs~\cite{Zanardi:2000zz,Zanardi:2001qu},
\be
\label{eq:ep_haar}
\ep(U) \equiv \avg{{\cal E}}\,.
\ee
where the overline denotes the average over Haar-random
product states $|\psi\> \otimes |\phi\>$. Equation~(\ref{eq:ep_haar})
is the original Zanardi--Zalka--Faoro convention and is also
used in Refs.~\cite{Beane:2018oxh,Hu:2025lua}. A
dimension-normalized convention, used in
Refs.~\cite{Low:2026evp,Low:2026kxb}, is related to it by
$\ep_{\rm N}=\frac{r}{r-1}\ep$, where
$r=\min(d_A,d_B)$ is the maximum Schmidt rank. Unless
otherwise stated, we use the convention in
Eq.~(\ref{eq:ep_haar}) throughout.
For two qubits, $\ep(U)=0$ precisely when $U$ is locally
equivalent to either the Identity or SWAP
gate~\cite{Low:2021ufv}.
By construction, $\ep(U)$ is a property of the
unitary operator itself, not of any particular input
state; it is also closely related to information
scrambling and operator
entanglement~\cite{Styliaris:2021prl}.

The Haar integrals in Eq.~(\ref{eq:ep_haar}) can be
evaluated analytically by mapping the problem to the
doubled Hilbert space.
The key idea is the operator--state correspondence: any
operator $O$ acting on a Hilbert space ${\cal H}$ of
dimension $d$ can be mapped to a state
$|O\>\!\> \in {\cal H} \otimes {\cal H}$ via
$|O\>\!\> = (O \otimes \Id)\sum_{i=1}^{d}|i\>|i\>$,
where $\{|i\>\}$ is an orthonormal basis of
${\cal H}$.
The Hilbert--Schmidt inner product of two operators
then becomes the overlap of their corresponding states:
$\<\!\< A | B \>\!\> = \tr(A^\dagger B)$.
Applying this correspondence to $U^{\otimes 2}$
acting on
$({\cal H}_A \otimes {\cal H}_B)^{\otimes 2}$,
the Haar averages reduce to traces of swap operators in
the doubled
space~\cite{Zanardi:2000zz,Zanardi:2001qu}.
Following the formulation
of Ref.~\cite{Lu:2008zz}, the result can be expressed
compactly as
\be
\label{eq:ep_def}
\ep(U) = 1 - C_{d_A}\,C_{d_B}
\sum_{\alpha=0}^{1} I_\alpha(U)\,,
\ee
where $C_d = 1/[d(d+1)]$ and, for $\alpha = 0, 1$,
\bea
\label{eq:Ialpha}
 I_\alpha(U) &=& \tr(T_{1+\alpha,\,3+\alpha})\nonumber\\
&& \quad
+ \< U^{\otimes 2},\;
T_{1+\alpha,\,3+\alpha} \cdot U^{\otimes 2}
\cdot T_{13}\>_{\text{HS}}\,.
\eea
Here $T_{ij}$ denotes the transposition (swap) operator
acting on the $i$th and $j$th tensor factors in
$({\cal H}_A \otimes {\cal H}_B)^{\otimes 2}$, and
$\< A, B\>_{\text{HS}} = \tr(A^\dagger B)$ is the
Hilbert--Schmidt inner product.
The advantage of this operator formulation is that
it reduces the computation of $\ep(U)$ to traces in the
doubled Hilbert space, avoiding any explicit
integration over random states.

\subsection{An Algorithm for the Time-average}
\label{subsect:time}

The time-evolution operator
$U(t) = e^{iHt}$ of a time-independent Hamiltonian
$H = \sum_n E_n |n\>\<n|$ generates a one-parameter family of
unitaries, and we define the time-averaged entanglement
power
\be
\label{eq:ep_avg}
\avg{\ep} \equiv \lim_{T\to\infty}
\frac{1}{T}\int_0^T \ep\bigl(e^{iHt}\bigr)\,dt\,,
\ee
which depends only on the spectrum and eigenstates
of the Hamiltonian $H$.

Although Eq.~(\ref{eq:ep_avg}) can be calculated by
brute-force, we present in the following an efficient
algorithm for its computation. Since $\ep(U(t))$ is quartic in the matrix elements of
$U(t)$, the integrand involves oscillatory phases
$e^{-i\omega\,t}$ where $\omega = E_m - E_n + E_p - E_q$.
Since
\be
\label{eq:time_avg_delta}
\lim_{T\to\infty}
\frac{1}{T}\int_0^T e^{-i\omega\,t}\,dt
= \delta_{\omega,\,0}\,,
\ee
the infinite-time average retains only those combinations
satisfying $\omega = 0$---the
diagonal-ensemble contribution. This enables an \emph{exact} evaluation of
$\avg{\ep}$ for any system whose Hamiltonian can be
diagonalized, without resorting to a finite integration
window. We outline the steps of this algorithm, leaving the full details in
Appendix~\ref{app:algorithm}:
\begin{enumerate}
\item Diagonalize $H$ to obtain eigenvalues $\{E_n\}$ and
eigenvectors $\{|n\>\}$.
\item For each eigenvector $|n\>$, expand it in the
product basis $\{|a\> \otimes |b\>\}$, where
$\{|a\>\}$ and $\{|b\>\}$ are orthonormal bases for
${\cal H}_A$ and ${\cal H}_B$ respectively, and arrange
the coefficients into a $d_A \times d_B$ matrix~$C_n$
with entries $(C_n)_{ab} = (\<a| \otimes \<b|)|n\>$.
\item The diagonal-ensemble condition
$E_m - E_n + E_p - E_q = 0$ is equivalent to
$E_m - E_p = E_n - E_q$, i.e.\ two pairs of
eigenstates contribute only when they share a common
eigenvalue difference
$\omega \equiv E_m - E_p = E_n - E_q$.
Collect all pairs $(k,l)$ with the same value of
$\omega = E_k - E_l$ into a group~$g_\omega$;
only pairs within the same group contribute to
$\avg{\ep}$.
\item For each group~$g_\omega$ containing $N_g$ pairs, form
the $d_A \times d_A$ matrix $M_i \equiv C_k C_l^\dagger$
for the $i$th pair, where $k$ and $l$ are the
eigenvector indices of that pair.
The contribution of this group to the time-averaged
purity is
$\sum_{i,j=1}^{N_g}|\tr(M_i^\dagger M_j)|^2$. 
\item Sum over all groups to obtain the dynamical
part of $\avg{I_0}$.
For $\avg{I_1}$, form also the $d_B \times d_B$
matrices $\hat{M}_i = C_k^\dagger C_l$ for each
pair; the contribution of each group is then
$\sum_{i,j=1}^{N_g}\tr(M_i^\dagger M_j)\cdot
\tr(\hat{M}_i^\dagger \hat{M}_j)$.
Adding the constant terms
$\tr(T_{13})$ and $\tr(T_{24})$ from
Eq.~(\ref{eq:Ialpha}), the time-averaged
entangling power $\avg{\ep}$ follows via
Eq.~(\ref{eq:ep_def}).
\end{enumerate}
The procedure requires only a single diagonalization and
linear algebra on $d_A \times d_B$ matrices, and is
practical for any system size where exact diagonalization
is feasible.  All time-averaged results in this work are
computed by this method.

It is worth pointing out that the time-averaged entanglement
power is determined by two ingredients: the eigenvalue
spectrum and the eigenvector structure.
The spectrum controls the grouping: eigenvalue degeneracies
enlarge the groups~$g_\omega$, increasing the number of
pairs that contribute to the sums in steps~4 and~5.
The eigenvectors, through the coefficient matrices~$C_n$,
determine the magnitude of each trace factor
$\tr(M_i^\dagger M_j)$ and
$\tr(\hat{M}_i^\dagger \hat{M}_j)$.
More symmetry typically produces more degeneracies,
increasing the time-averaged purity
$\avg{I_0 + I_1}$ and thereby suppressing $\avg{\ep}$.

\section{Two-Site Spin Chain}
\label{sec:twosite}

We begin by investigating the entangling power in the
simplest nontrivial setting: the two-site anisotropic
Heisenberg Hamiltonian
\be
\label{eq:H_2site}
H = a_x\, S_x\!\otimes\! S_x
  + a_y\, S_y\!\otimes\! S_y
  + a_z\, S_z\!\otimes\! S_z\,,
\ee
where $S_i$ are the spin-$s$ angular momentum operators
acting on a $d$-dimensional Hilbert space at each
site, with $d = 2s + 1$, and $a_j$'s are exchange constants.
The time-evolution operator is $U(t) = e^{iHt}$.
From the perspective of quantum information, the
two-site spin chain is a bipartite quantum gate:
spin-1/2 ($d = 2$) corresponds to a two-qubit
unitary, spin-1 ($d = 3$) to a two-qutrit unitary,
and general spin-$s$ to a two-qudit system with
local dimension~$d$.
This mapping places the entangling power of
spin-chain Hamiltonians in direct contact with
the theory of entangling gates in quantum computation
and allows the results of this section to be read
in either language.

The special cases of Eq.~(\ref{eq:H_2site}) include the
isotropic XXX model ($a_x = a_y = a_z$), the
XXZ model ($a_x = a_y$, $a_z = \Delta$), the XX model
($a_x = a_y$, $a_z = 0$), and the Ising model
($a_x = a_y = 0$, $a_z \neq 0$), among others.
The two-site model is exactly solvable for arbitrary
spin, and the analytical results presented below provide
the intuition for the finite-size and thermodynamic-limit
analyses in subsequent sections.

\subsection{Spin-1/2 (Two-qubit)}
\label{sec:spin_half}

For spin-1/2, the single-site operators are
$S_i = \sigma_i/2$ with $\sigma_i$ the Pauli matrices.
The two-site Hilbert space has dimension $d_1 d_2 = 4$,
the doubled space has dimension~16, and the normalization
constant is $C_2 = 1/6$.
A direct  computation yields the closed-form
entangling power
\bea
\label{eq:ep_d2_freq}
\ep(U) &=& \frac{1}{36}\Big(6
- \cos\!\big[(a_x\!-\!a_y)t\big] - \cos\!\big[(a_x\!+\!a_y)t\big]
\non\\
&&- \cos\!\big[(a_x\!-\!a_z)t\big]
- \cos\!\big[(a_y\!-\!a_z)t\big]
\non\\
&&- \cos\!\big[(a_x\!+\!a_z)t\big] - \cos\!\big[(a_y\!+\!a_z)t\big]\Big)\,,
\eea
which involves six  ``frequencies''
$a_x \pm a_y$, $a_x \pm a_z$, $a_y \pm a_z$.
For the special cases of interest, Eq.~(\ref{eq:ep_d2_freq}) reduces to
\bea
\label{eq:ep_d2_special}
{\rm XXX:} && \ep = \tfrac{1}{6}\sin^2(at)\,,
\non\\[4pt]
{\rm Ising:} && \ep = \tfrac{2}{9}\sin^2\!\big({at}/{2}\big)\,,
\non\\[4pt]
{\rm XX:} && \ep = \tfrac{1}{9}\big(3+\cos(at)\big)\sin^2\!\big({at}/{2}\big)\,.
\eea
The XXX result $\ep = \frac{1}{6}\sin^2(at)$
has a transparent gate-theoretic interpretation.
Using $\vec{\sigma}_1 \!\cdot\! \vec{\sigma}_2 = 2\,{\rm SWAP} - I$,
the time-evolution operator can be written as
\be
U(t) = e^{i\alpha(t)}\big[\cos(at/2)\,I
- i\sin(at/2)\,{\rm SWAP}\big]\ ,
\ee
where $\alpha(t)$ is an overall phase.
At $at = 0$ one has $U = I$ (the identity gate), while
at $at = \pi$ one has $U \propto {\rm SWAP}$;
these are the only two
two-qubit gates with vanishing entanglement
power~\cite{Low:2021ufv}.
The $\sin^2(at)$ oscillation thus traces the entanglement
power of $U(t)$ as it interpolates between these two
zero-entanglement gates.

\begin{figure*}[t]
\centering
\includegraphics[width=\textwidth]{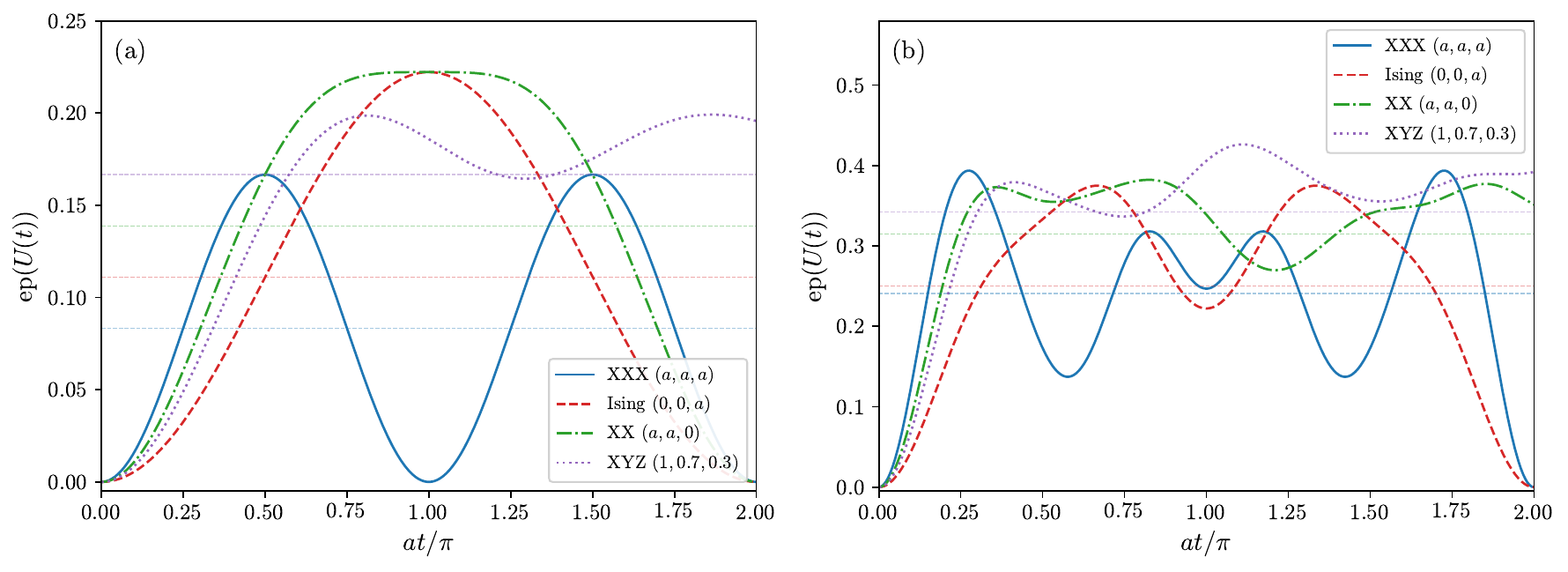}
\caption{\em Instantaneous ep $\ep(U(t))$ of
the two-site Hamiltonian~(\ref{eq:H_2site}) as a function
of time $at/\pi$ for the XXX (solid), Ising
(dashed), XX (dash-dotted), and XYZ (dotted) models.
(a) Spin-1/2 (two-qubit, $d = 2$).
(b) Spin-1 (two-qutrit, $d = 3$).
For the XYZ model we use
$(a_x, a_y, a_z) = (1, 0.7, 0.3)$.
Thin dashed horizontal lines indicate the time-averaged
values $\avg{\ep}$.
}
\label{fig:ep_instantaneous}
\end{figure*}

Figure~\ref{fig:ep_instantaneous}(a) shows
$\ep(U(t))$ as a function of time for the four models.
The XXX model exhibits a clean single-frequency
oscillation $\ep \propto \sin^2(at)$, while the Ising
model has a slightly richer pattern with period
doubled.
The XX model shows additional beating between its two
distinct nonzero frequencies $a$ and $2a$
(the remaining four of the six cosines in
Eq.~(\ref{eq:ep_d2_freq}) collapse onto these two values
when $a_x = a_y = a$, $a_z = 0$),
and the XYZ model, with all six
frequencies active, produces the most complex
oscillatory structure.
Despite this diversity in the time dependence, the time
averages (dashed lines) respect the symmetry hierarchy
in every case.

The infinite-time average of $\cos(\omega t)$ vanishes for
$\omega \neq 0$ and equals~1 for $\omega = 0$.
From Eq.~(\ref{eq:ep_d2_freq}), the time-averaged
entangling power is therefore
\be
\label{eq:ep_avg_d2}
\avg{\ep} = \frac{6 - N_0}{36}\,,
\ee
where $N_0$ is the number of the six frequencies that vanish.
Greater symmetry among the couplings forces more frequency
differences to zero, directly reducing $\avg{\ep}$.
Table~\ref{tab:ep_d2} lists $\avg{\ep}$ for each model,
and the resulting hierarchy is
\be
\label{eq:hierarchy_d2}
\avg{\ep}_{\rm XYZ} > \avg{\ep}_{\rm XX}
= \avg{\ep}_{\rm XXZ}
> \avg{\ep}_{\rm Ising}
= \avg{\ep}_{\rm XX}
> \avg{\ep}_{\rm XXX}\,.
\ee

\begin{table}[t]
\centering
\begin{tabular}{l|c|c|c|c|c}
\toprule
Model & $(a_x, a_y, a_z)$ & Symmetry & $N_0$ & $\avg{\ep}$ & Decimal \\
\hline
\midrule
XYZ & all distinct & --- & 0 & $1/6$ & 0.1667 \\
XX & $(a,a,0)$ & $U(1)$ & 1 & $5/36$ & 0.1389 \\
XXZ (generic) & $(a,a,\Delta)$ & $U(1)$ & 1 & $5/36$ & 0.1389 \\
Ising & $(0,0,a)$ & $U(1)^2$ & 2 & $1/9$ & 0.1111 \\
XXX & $(a,a,a)$ & $SU(2)$ & 3 & $1/12$ & 0.0833 \\
\bottomrule
\end{tabular}
\caption{\em Time-averaged entangling power for two-qubit
models and their continuous symmetry groups.
}
\label{tab:ep_d2}
\end{table}

To understand the ordering of $\avg{\ep}$ values in
Table~\ref{tab:ep_d2}, it is helpful to identify the
continuous symmetries of each model.
A direct computation of the commutators
$[S_i^{\rm tot}, H]$ yields
\be
\label{eq:commutator}
[S_z^{\rm tot},\, H] = i(a_x - a_y)
\big(S_{y,1}S_{x,2} + S_{x,1}S_{y,2}\big)\,,
\ee
and cyclic permutations for $S_x^{\rm tot}$ and
$S_y^{\rm tot}$.
Thus $S_z^{\rm tot}$ is conserved if and only if (iff)
$a_x = a_y$, and similarly $S_x^{\rm tot}$ iff
$a_y = a_z$, and $S_y^{\rm tot}$ iff $a_z = a_x$.
The XYZ model, with all couplings distinct, has no
continuous symmetry.
The XXZ and XX models, with $a_x = a_y$, conserve
$S_z^{\rm tot}$ and thus have $U(1)$ symmetry.
The Ising model $H = a\,S_z \!\otimes\! S_z$ commutes
with $S_{z,1}$ and $S_{z,2}$ independently---the spin
projection at each site is separately conserved---giving
$U(1) \!\times\! U(1)$, which has {\em more} continuous
symmetry than XXZ or XX despite involving only one
nonzero coupling.
At the XXX point all three components of
$\vec{S}^{\,\rm tot} = \vec{S}_1 + \vec{S}_2$ are
conserved, promoting the symmetry to $SU(2)$.
The resulting symmetry hierarchy is
none (XYZ) $\subset$ $U(1)$ (XXZ/XX) $\subset$
$U(1)\!\times\!U(1)$ (Ising) $\subset$ $SU(2)$
(XXX), with dimensions $0, 1, 2, 3$.
As the dimension grows, the number of vanishing
frequencies $N_0$ increases in lockstep: each new
conserved charge enforces additional degeneracies in
the eigenvalue spectrum, which in turn forces more
oscillatory terms to average to zero, directly
suppressing $\avg{\ep}$.

\subsection{Spin-1 (Two-qutrit)}
\label{sec:spin_one}

For spin-1, the single-site operators are the standard
$3 \times 3$ angular momentum matrices.
The two-site Hilbert space has dimension
$d_1 d_2 = 9$, the doubled space in Eq.~(\ref{eq:Ialpha}) has dimension~81,
and $C_3 = 1/12$.
The transposition operators $T_{13}$ and $T_{24}$ are
$81 \times 81$ permutation matrices with
$\tr(T_{13}) = \tr(T_{24}) = 27$.
The eigenvalue structure is considerably richer than
the spin-1/2 case, as we will see.

Analytic computation yields closed-form expressions for
the XXX, Ising, and XXZ models.
For the general XYZ model the time averages are
obtained by numerically diagonalizing $H$ and then
performing the exact infinite-time average using the
eigenvalue-grouping algorithm described in
Sec.~\ref{subsect:time}.

\paragraph{XXX ($a_x = a_y = a_z = a$).}
\bea
\label{eq:ep_d3_heis}
\ep &=& \frac{1}{648}\Big(156 - 15\cos(at)
- 6\cos(2at)
\non\\
&& - 65\cos(3at) - 60\cos(4at)
- 10\cos(6at)\Big)\,.
\eea
This involves harmonics at frequencies
$a, 2a, 3a, 4a, 6a$, considerably richer than
the single-frequency spin-1/2 result
$\ep = ({1}/{6})\sin^2(at)$.

\paragraph{Ising / XX ($a_x = a_y = 0$, $a_z = a$).}
\be
\label{eq:ep_d3_ising}
\ep = \frac{1}{36}\Big(9 - 4\cos(at)
- 4\cos(2at) - \cos(4at)\Big)\,.
\ee
By relabeling of spin components, this formula
applies to all single-coupling models (XX, YY, ZZ).

\paragraph{XX ($a_x = a_y = a$, $a_z = 0$).}
\bea
\label{eq:ep_d3_xy}
\ep &=& \frac{1}{2304}\Big(725
- 128\cos(2at) - 48\cos(4at)
\non\\
&& - 192\cos(\sqrt{2}\,at)
- 92\cos(2\sqrt{2}\,at)
\non\\
&& - 9\cos(4\sqrt{2}\,at)
- 96\cos\!\big[(\sqrt{2}\!-\!2)at\big]
\non\\
&& - 32\cos\!\big[2(\sqrt{2}\!-\!1)at\big]
- 32\cos\!\big[2(\sqrt{2}\!+\!1)at\big]
\non\\
&& - 96\cos\!\big[(\sqrt{2}\!+\!2)at\big]\Big)\,.
\eea
The appearance of irrational frequencies
($\sqrt{2}\,a$, etc.)\ reflects the eigenvalue structure
of the combined
$S_x \!\otimes\! S_x + S_y \!\otimes\! S_y$ Hamiltonian.

\paragraph{XXZ ($a_x = a_y = 1$, $a_z = \Delta$).}
The instantaneous entangling power takes the form of
a trigonometric polynomial with 24 frequencies:
\be
\label{eq:ep_d3_xxz}
\ep(t)
= \frac{5}{8}
- \frac{1}{144}\bigg[
\mathcal{A}_0
+ \sum_{j=1}^{24} \mathcal{A}_j\,\cos(\Omega_j\, t)
\bigg]\,,
\ee
where the DC component is
$\mathcal{A}_0
= (46\Delta^4 + 684\Delta^2 + 2636)/(\Delta^2+8)^2$.
Of the 24 frequencies, 8 are rational:
\bea
\label{eq:ep_xxz_rational}
&&\hspace{-1cm}
\sum_{j\in R}\!
\mathcal{A}_j \cos(\Omega_j t)
\;=\;
\frac{4}{\Delta^2\!+\!8}\cos(\Delta\, t)
\non\\
&&
+\;\frac{2\Delta^2}{\Delta^2\!+\!8}
\Big[\cos\!\big((\Delta\!+\!2)t\big)
 + \cos\!\big((\Delta\!-\!2)t\big)\Big]
\non\\
&&
+\; 4\Big[\cos\!\big((2\Delta\!+\!2)t\big)
 + \cos\!\big((2\Delta\!-\!2)t\big)\Big]
\non\\
&&
+\; \frac{8}{\Delta^2\!+\!8}\cos(3\Delta\, t)
+ 2\cos(4\Delta\, t)
\non\\
&&
+\; 3\cos(4t)\,,
\eea
and the remaining 16 involve the irrational quantity
$\zeta \equiv \sqrt{\Delta^2 + 8}$ through
frequencies such as $\varsigma$,
$(3\Delta \pm \varsigma)/2$,
$(\Delta \pm \varsigma)/2 \pm 2$, etc.
These are tabulated in full in
Table~\ref{tab:fourier_xxz}; the complete derivation
of both the instantaneous and time-averaged formulas
is given in Appendix~\ref{app:xxz_analytic}.

Time averaging kills all oscillating terms,
leaving only the DC component.  After simplification,
the irrational eigenvalues
$E_\pm = (-\Delta \pm \varsigma)/2$
cancel pairwise, yielding the closed-form result
\be
\label{eq:ep_xxz}
\avg{\ep}(\Delta)
= \frac{11\Delta^4 + 189\Delta^2 + 781}
       {36(\Delta^2 + 8)^2}\,,
\ee
valid for generic $\Delta$.

Fig.~\ref{fig:ep_instantaneous}(b)
shows the instantaneous entangling power for the
spin-1 models.
The richer harmonic content compared to spin-1/2 is
 apparent: the XXX model now involves
five frequencies ($a, 2a, 3a, 4a, 6a$) rather than a
single one, and the XX model exhibits irrational-frequency
beating that never recurs exactly.
Nevertheless, the time averages (dashed lines) again
respect the symmetry hierarchy.

The time-averaged values, obtained from the constant
terms in the above expressions, are collected in
Table~\ref{tab:ep_d3}.
The hierarchy is qualitatively the same as for spin-1/2:
\be
\label{eq:hierarchy_d3}
\avg{\ep}_{\rm XYZ} > \avg{\ep}_{\rm XXZ}
> \avg{\ep}_{\rm XX}
> \avg{\ep}_{\rm Ising}
= \avg{\ep}_{\rm XX}
> \avg{\ep}_{\rm XXX}\,.
\ee
It is worth noting that, unlike spin-1/2, the XX and
generic XXZ values now differ:
the XX model gives $\avg{\ep} = 725/2304 \approx 0.315$,
while Eq.~(\ref{eq:ep_xxz}) evaluated at $\Delta = 0$
gives $781/2304 \approx 0.339$.
The degeneracy
$\avg{\ep}_{\rm XX} = \avg{\ep}_{\rm XXZ}$ seen in
the $d = 2$ case is thus broken for spin-1;
the origin of this splitting will become clear in the
eigenvalue analysis of Sec.~\ref{sec:eigenvalue}.

\begin{table}[t]
\centering
\begin{tabular}{l|c|c|c|cc}
\toprule
Model & $(a_x, a_y, a_z)$ & Symmetry & $\avg{\ep}$ (exact) & Decimal \\
\midrule
\hline
XYZ & $(1,\, 0.5,\, 0.25)$ & --- & --- & 0.371 \\
XYZ & $(1,\, 0.7,\, 0.3)$ & --- & --- & 0.377 \\
XXZ & $(1,\, 1,\, 0.5)$ & $U(1)$ & Eq.~(\ref{eq:ep_xxz}) & 0.338 \\
XX & $(a,\, a,\, 0)$ & $U(1) \times \mathbb{Z}_2$ & $725/2304$ & 0.315 \\
Ising & $(0,\, 0,\, a)$ & $U(1)^2$ & $1/4$ & 0.250 \\
XXX & $(a,\, a,\, a)$ & $SU(2)$ & $13/54$ & 0.241 \\
\bottomrule
\end{tabular}
\caption{{\em Time-averaged entangling power for
spin-1 models and their continuous symmetry groups.
}}
\label{tab:ep_d3}
\end{table}

\begin{figure}[t]
\centering
\includegraphics[width=\columnwidth]{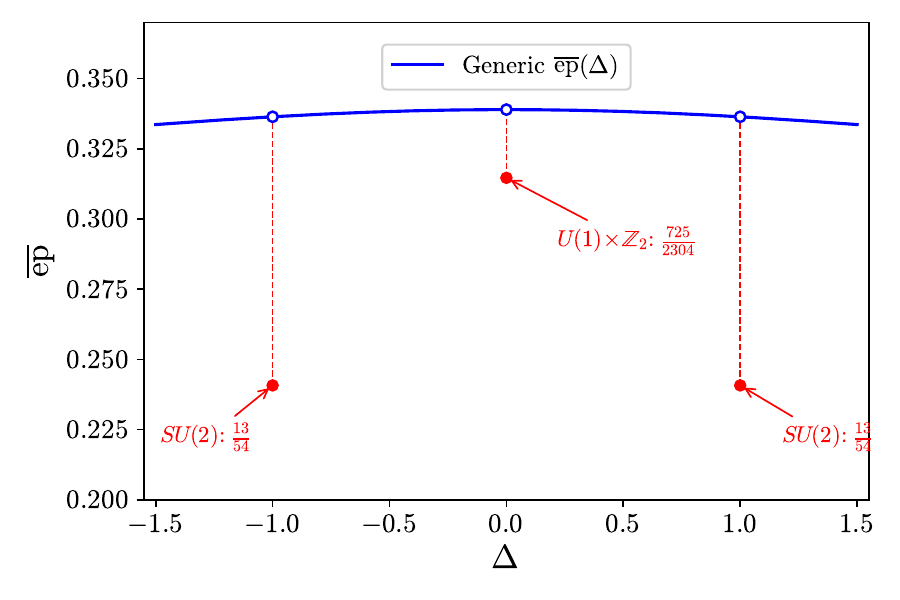}
\caption{{\em Time-averaged entangling power of the
two-site spin-1 XXZ model as a function of anisotropy
$\Delta$.
The solid curve is the generic analytic formula,
Eq.~(\ref{eq:ep_xxz}).
At $\Delta = \pm 1$ ($SU(2)$ symmetry) and
$\Delta = 0$ ($U(1) \times \mathbb{Z}_2$ symmetry),
eigenvalue degeneracies result in 
the true $\avg{\ep}$ (filled circles) dropping below
the generic curve (open circles). Dashed lines indicate the magnitude of each dip.
}}
\label{fig:ep_xxz_analytic}
\end{figure}

The analytic formula Eq.~(\ref{eq:ep_xxz}) and
the degenerate-point values are displayed in
Fig.~\ref{fig:ep_xxz_analytic}.
The generic curve is nearly flat, with a shallow
minimum at $\Delta = 0$ (where
$\avg{\ep} = 781/2304 \approx 0.339$) and a
plateau at $11/36 \approx 0.306$ for
$|\Delta| \to \infty$.
Superimposed on this smooth background are sharp
dips at $\Delta = \pm 1$ (where $SU(2)$ symmetry
introduces additional eigenvalue degeneracies, giving
$\avg{\ep} = 13/54$) and a milder dip at
$\Delta = 0$ (where a $\mathbb{Z}_2$
spectrum-flipping symmetry forces further
degeneracies, giving $\avg{\ep} = 725/2304$).
We now identify the symmetry at each of these
special points.

\paragraph{$\Delta = \pm 1$: $SU(2)$ symmetry.}
At $\Delta = 1$ the XXZ Hamiltonian becomes the
isotropic XXX model, with full $SU(2)$ symmetry.
The dip at $\Delta = -1$ is exactly as deep,
$\avg{\ep}(-1) = \avg{\ep}(1)$, as a consequence of
local unitary equivalence.
Define $\pi$-rotations about the $x$ and $y$ axes,
$W_x = e^{i\pi S_x}$ and $W_y = e^{i\pi S_y}$, which
act on the spin operators as
\bea
\label{eq:LU_rotations}
W_x &:& (S_x, S_y, S_z) \to (S_x, -S_y, -S_z)\,,
\non\\
W_y &:& (S_x, S_y, S_z) \to (-S_x, S_y, -S_z)\,.
\eea
Under the local unitary $W_x \!\otimes\! W_y$, the
Hamiltonian transforms as
\bea
\label{eq:LU_transform}
&& (W_x\!\otimes\! W_y)\, H(a_x, a_y, a_z)\,
(W_x\!\otimes\! W_y)^\dagger\nonumber\\
&&\qquad \qquad \qquad \qquad \quad= H(-a_x, -a_y, a_z)\,.
\eea
Applied to the XXZ point $(1,1,-\Delta)$, this gives
$H(-1,-1,-\Delta) = -H(1,1,\Delta)$.
Since the entangling power satisfies
$\ep(e^{-iHt}) = \ep(e^{iHt})$,
the time averages are identical:
$\avg{\ep}(1,1,-\Delta) = \avg{\ep}(1,1,\Delta)$.
Note that this argument is valid for any spin $d$,
not just spin-1.

\paragraph{$\Delta = 0$: $\mathbb{Z}_2$
spectrum-flipping symmetry.}
At $\Delta = 0$ the model reduces to
$H = S_x \!\otimes\! S_x + S_y \!\otimes\! S_y$
(the XX model).
This Hamiltonian possesses a discrete $\mathbb{Z}_2$
symmetry generated by $W = e^{i\pi S_{z,1}}$ acting on
site~1 alone, which satisfies $W S_z W^\dagger = S_z$
but $W S_\pm W^\dagger = -S_\pm$.
This implies
$(W\!\otimes\! I)\,H\,(W\!\otimes\! I)^\dagger = -H$,
so the spectrum is symmetric about zero: for every
eigenvalue $E$, $-E$ is also an eigenvalue with the
same degeneracy.
This spectrum-flipping symmetry forces additional
eigenvalue degeneracies that suppress $\avg{\ep}$
below the generic XXZ value.
It is worth noting that, for spin-1/2, $\Delta = 0$ is
also the free-fermion point via the Jordan--Wigner
transformation; this connection will play an important
role in Sec.~\ref{sec:ep_smatrix}.

\subsection{Eigenvalue analysis}
\label{sec:eigenvalue}

The mechanism behind the correlation between symmetry
and entanglement suppression can be understood from the
spectral decomposition of the time-evolution operator.
Writing $U(t) = \sum_k e^{iE_k t}\, P_k$, where $P_k$
is the projector onto the eigenspace of eigenvalue $E_k$,
and recalling from Sec.~\ref{subsect:time} that the
infinite-time average retains only terms satisfying the
diagonal-ensemble condition
$E_m - E_n + E_p - E_q = 0$, the time-averaged
entangling power depends in general on two ingredients:
the eigenvalue spectrum (which controls which terms
survive the time average) and the eigenvectors (which
determine the matrix elements of those surviving terms).
Recall that the time-averaging algorithm of
Sec.~\ref{subsect:time} partitions eigenstate pairs
into $\omega$-groups sharing the same energy difference.
When a symmetry enforces eigenvalue degeneracies,
pairs that would belong to separate groups at generic
coupling are merged into larger groups, and the coherent
sum within each enlarged group enhances the purity sums,
thereby suppressing $\avg{\ep}$.
As we now discuss, the spin-1/2 case is special in that
the eigenvectors are coupling-independent, so $\avg{\ep}$
is determined entirely by the eigenvalue spectrum.
For spin-1, both ingredients play a role.

\paragraph{Spin-1/2.}
The Hamiltonian~(\ref{eq:H_2site}) has eigenvalues
\bea
\label{eq:eigenvalues_d2}
E \in \bigg\{
&&\!\!\!\!\!\!\!\frac{-a_x \!-\! a_y \!-\! a_z}{4},\;\;
\frac{a_x \!+\! a_y \!-\! a_z}{4},
\non\\
&&\frac{a_x \!-\! a_y \!+\! a_z}{4},\;\;
\frac{-a_x \!+\! a_y \!+\! a_z}{4}\bigg\},
\eea
and, crucially, the eigenvectors are the four Bell
states {\em independent} of the coupling constants
$a_x, a_y, a_z$.
Because the projectors $P_k$ are universal, the
time-averaged entangling power depends {\em only}
on the eigenvalue spectrum.
Moreover, the closed-form expression
Eq.~(\ref{eq:ep_d2_freq}) shows that this dependence
takes a particularly simple form: six cosine terms,
each with equal weight $1/36$, giving
$\avg{\ep} = (6 - N_0)/36$
[Eq.~(\ref{eq:ep_avg_d2})], where $N_0$ is the number
of the six frequencies $a_x \pm a_y$, $a_x \pm a_z$,
$a_y \pm a_z$ that vanish.
The quantity $N_0$ therefore determines $\avg{\ep}$
completely, and models sharing the same $N_0$ necessarily
share the same $\avg{\ep}$---this is why the XXZ and XX
models, despite having different eigenvalue structures,
both give $\avg{\ep} = 5/36$: each has exactly one
vanishing frequency ($a_x - a_y = 0$), so $N_0 = 1$.
The values of $N_0$ for all models are already listed
in Table~\ref{tab:ep_d2}.

At the XXX point, the four eigenvalues
collapse to just two distinct values:
$E = -3a/4$ (singlet, $S = 0$) and
$E = a/4$ (triplet, $S = 1$).
The $SU(2)$ symmetry forces the degeneracy
$E_2 = E_3 = E_4$, which causes three of the six
frequencies to vanish simultaneously ($N_0 = 3$),
yielding the minimum $\avg{\ep} = 1/12$.

This equal-weight
simplicity---in which $N_0$ alone determines
$\avg{\ep}$---is specific to spin-1/2 and relies on
the universality of the Bell-state eigenvectors.
For spin-1, the eigenvectors depend on the coupling
constants and a more refined analysis is
needed.

\paragraph{Spin-1.}
The eigenvalues for the key models are:
XXX $\{-2a, -a, a\}$ with degeneracies
$\{1, 3, 5\}$ ($S = 0, 1, 2$);
Ising $\{-a, 0, a\}$ with degeneracies
$\{2, 5, 2\}$;
and XX $\{-\!\sqrt{2}\,a, -a, 0, a, \sqrt{2}\,a\}$ with
degeneracies $\{1, 2, 3, 2, 1\}$.

Unlike the spin-1/2 case, the eigenvectors of the
spin-1 Hamiltonian {\em do} depend on the coupling
constants.
At the XXX point, $SU(2)$ symmetry organizes
the 9 two-site states into irreducible representations
of total spin: $S = 0$,
$S = 1$, and $S = 2$, with eigenvalues $-2a$, $-a$, and $a$,
respectively.
Away from the isotropic point, the eigenstates are no
longer organized by total spin.

Following the algorithm of Sec.~\ref{subsect:time}, we
partition the $d^4 = 81$ ordered pairs $(k,l)$ of eigenstates
into $\omega$-groups labeled by $\omega = E_k - E_l$.
The diagonal-ensemble condition
$E_m - E_n = E_p - E_q$ is satisfied whenever
$(m,n)$ and $(p,q)$ belong to the same $\omega$-group.
The number of distinct $\omega$-groups for each model is:
XYZ~(61), XXZ~(27), XX~(13), XXX~(7), and
Ising~(5).
The XYZ and XXZ counts are generic values; specific
parameter choices can produce accidental degeneracies
that reduce the count.
Note that the XX model has the same eigenvalue structure
as Ising by rotational equivalence, and hence the same
$\avg{\ep} = 1/4$.

The ordering of $\avg{\ep}$ broadly tracks the
$\omega$-group count: more groups means more independent
oscillatory terms survive the time average, leading to a
smaller $\avg{I_0 + I_1}$ and hence a higher
$\avg{\ep}$.  However, the $\omega$-group count alone
does not determine $\avg{\ep}$ for spin-1, since the
eigenvectors---unlike the spin-1/2 case---depend on the
coupling constants and enter the time-averaged purities
through the matrix elements $\tr_A(M_i^\dagger M_j)$.
The XXX model, with 7 groups, achieves
$\avg{\ep} = 13/54 \approx 0.241$, lower than the
Ising model's $\avg{\ep} = 1/4 = 0.250$ with only 5
groups.  To disentangle the eigenvalue and eigenvector
contributions to this suppression, we turn to the
isospectral decomposition in the next subsection.

\subsection{Isospectral decomposition}
\label{sec:isospectral}

The eigenvalue analysis of the preceding subsection shows
that eigenvalue degeneracy counting alone cannot fully
explain the suppression of $\avg{\ep}$ at symmetric
points, particularly for spin-1.
To isolate the eigenvector contribution, we  construct
isospectral families---Hamiltonians sharing the same
spectrum as the XXX model but with different
eigenstates---and compute their entangling power.

\paragraph{Uniqueness of the XXX degeneracy.}
\label{sec:spin1_unique}
We first ask whether the XXX model is the
{\em only} member of the XYZ family in Eq.~(\ref{eq:H_2site})
with the XXX eigenvalue degeneracy pattern.
For spin-1/2, the eigenvalue degeneracy pattern
$\{1, 3\}$ requires all three triplet eigenvalues in
Eq.~(\ref{eq:eigenvalues_d2}) to coincide, which forces
$a_x = a_y = a_z$---precisely the XXX point.
Moreover, sign-flipping local unitaries of the form
$e^{i\pi S_\alpha} \!\otimes\! I$ map
$(a_x, a_y, a_z) \mapsto (\pm a_x, \pm a_y, \pm a_z)$
(with exactly two sign flips), so any XYZ point with
$|a_x| = |a_y| = |a_z|$ is locally unitarily equivalent
to the XXX model.
The $\{1, 3\}$ degeneracy pattern is therefore unique
to the XXX local-unitary equivalence class.

For spin-1, the general XYZ Hamiltonian does not commute
with $S_z^{\rm tot}$, but it does commute with the
$\mathbb{Z}_2 \times \mathbb{Z}_2$ parity operators
$R_\alpha \equiv e^{i\pi S_{\alpha,1}} \otimes
e^{i\pi S_{\alpha,2}}$ for $\alpha = x,y,z$.
Since a $\pi$-rotation about the $\alpha$-axis preserves
$S_\alpha$ and flips the other two components, each
bilinear term $S_{\beta,1} S_{\beta,2}$ picks up
two sign flips that cancel, giving
$[R_\alpha, H] = 0$.
As shown in Appendix~\ref{app:Z2Z2}, the
$9 \times 9$ Hilbert space decomposes under
$(R_z, R_x)$ into four sectors of dimension $3+2+2+2$.
The three two-dimensional sectors have eigenvalues
$\pm a_z$, $\pm a_x$, and $\pm a_y$, respectively,
while the three-dimensional sector has the following characteristic polynomial:
\be
\label{eq:depressed_cubic}
p(\lambda) = \lambda^3
- (a_x^2 + a_y^2 + a_z^2)\,\lambda
+ 2\,a_x a_y a_z\,.
\ee
For the XXX degeneracy pattern $\{1, 3, 5\}$,
the 5-fold degenerate eigenvalue must appear in all four
sectors, which forces $|a_x| = |a_y| = |a_z|$.
Consistency with Eq.~(\ref{eq:depressed_cubic}) is confirmed by the identity
$p(a_x) = -a_x(a_y - a_z)^2$,
so that $a_x$ is a root of $p$ if and only if $a_y = a_z$.
Cyclic permutations then force
$a_x = a_y = a_z$---again the XXX point
(up to local unitary equivalence via sign flips).

In fact, the $\mathbb{Z}_2\times\mathbb{Z}_2$ decomposition
derived in Appendix~\ref{app:Z2Z2} makes it possible to
compute $\avg{\ep}$ for the XXZ model in Eq.~(\ref{eq:ep_xxz});
the detail is given in
Appendix~\ref{app:xxz_analytic}.

\paragraph{Isospectral family on $\CC\mathbb{P}^3$.} In the remainder of this subsection
 we restrict to
spin-1/2 where the two-site Hilbert space
$\CC^4$ is simple and straightforward to analyze.
Having established that the XXX degeneracy pattern
is unique within the XYZ family, we now consider
Hamiltonians {\em outside} this family that share the
XXX spectrum
$\{-3J/4, J/4, J/4, J/4\}$ but have different eigenstates. The eigenstate for $-3J/4$ can be parameterized by a normalized vector $|z\> \in \CC\mathbb{P}^3$ with the corresponding projector 
$|z\>\<z|$.
The Hamiltonian is then
\be
\label{eq:H_iso}
H = \frac{J}{4}\,\Id - J\,|z\>\<z|
\ee
which satisfies $H|z\>=(-3J/4)|z\>$. The rank-3 projector $\Id-|z\>\<z|$ then spans  the eigenspace with eigenvalue $J/4$.
The isospectral family is thus parametrized by
$\CC\mathbb{P}^3 = U(4)/(U(3) \times U(1))$,
a 6-dimensional manifold.
The XXX model corresponds to $|z\>$ being
the maximally entangled Bell state $(|01\> - |10\>)/\sqrt{2}$.

\paragraph{Spin-operator representation.}
To make explicit the physical content of the isospectral
Hamiltonian in Eq.~(\ref{eq:H_iso}), we expand the projector
$|z\>\<z|$ in the operator basis
$\{\Id, \sigma_i\} \otimes \{\Id, \sigma_j\}$.
Writing $|z\> = (z_1, z_2, z_3, z_4)$ in the
computational basis with $\sum_k |z_k|^2 = 1$,
the completeness of the Pauli basis gives
\bea
|z\>\<z| &=& \frac{1}{4}\Id
+ \frac{1}{2}\sum_i \<\sigma_i\>_1\, S_{i,1}
+ \frac{1}{2}\sum_i \<\sigma_i\>_2\, S_{i,2}
\non\\
&& + \sum_{i,j} \<\sigma_i\!\otimes\!\sigma_j\>\,
S_{i,1}\,S_{j,2}\,,
\eea
where $\<\cdots\>$ denotes the expectation value in
$|z\>$.
The identity piece cancels the $\frac{J}{4}\,\Id$ in
Eq.~(\ref{eq:H_iso}), leaving
\bea
\label{eq:H_spin_op}
H &=& -\frac{J}{2}\sum_i
\Big(\<\sigma_i\>_1\, S_{i,1}
+ \<\sigma_i\>_2\, S_{i,2}\Big)
\non\\
&& - J\sum_{i,j}
\<\sigma_i\!\otimes\!\sigma_j\>\,
S_{i,1}\,S_{j,2}\,.
\eea
The most general isospectral Hamiltonian is therefore a
bilinear spin--spin coupling with single-site Zeeman fields.
All 15 parameters (9 couplings and 6 field components)
are fixed by the 6 real parameters of
$|z\> \in \CC\mathbb{P}^3$.

It is instructive to verify Eq.~(\ref{eq:H_spin_op})
in two limiting cases.
For the XXX model,
$|z\> = (|01\> - |10\>)/\sqrt{2}$ (the singlet), all
single-site expectations vanish and
$\<\sigma_i\!\otimes\!\sigma_j\> = -\delta_{ij}$, so
that Eq.~(\ref{eq:H_spin_op}) reduces to
$H = J\,\vec{S}_1 \cdot \vec{S}_2$ as expected.
At the opposite extreme, taking $|z\> = |00\>$ (an unentangled
product state) gives
$\<\sigma_z\>_1 = \<\sigma_z\>_2 = 1$ as the only
nonvanishing single-site expectations and
$\<\sigma_z\!\otimes\!\sigma_z\> = 1$ as the only
nonvanishing correlator.
Equation~(\ref{eq:H_spin_op}) then yields
the product-state partner
\be
\label{eq:H_prod}
H_{\rm prod} = -\frac{J}{2}\big(S_{z,1} + S_{z,2}\big)
- J\,S_{z,1} S_{z,2}\,,
\ee
an Ising interaction plus a longitudinal magnetic field,
whose eigenstates
$\{|00\>, |01\>, |10\>, |11\>\}$ are all product states
with eigenvalues
$\{-3J/4, J/4, J/4, J/4\}$, identical to the
XXX model.
Despite having the same spectrum, $H_{\rm prod}$ in Eq.~(\ref{eq:H_prod})
gives
$\avg{\ep}_{\rm prod} = 1/9 \approx 0.1111$,
which is {\em larger} than the XXX value
$\avg{\ep}_{\rm XXX} = 1/12 \approx 0.0833$ by a
factor of $4/3$.
The gap between the two is purely an eigenvector effect.

\paragraph{Local-unitary equivalence and the entangling power.}
The XXX point and any other point in the isospectral
family are related by a unitary transformation,
$H(|z\>) = V\,H_{\rm XXX}\,V^\dagger$ with $V \in SU(4)$.
Such a transformation preserves the spectrum but in general
changes $\avg{\ep}$: the entangling power is invariant only
under the subgroup of single-qubit (local) unitaries
$SU(2) \otimes SU(2) \subset SU(4)$.
A generic $V \in SU(4)$ therefore generates a Hamiltonian
with the same eigenvalues as $H_{\rm XXX}$ but a distinct
entangling power.

\begin{figure}[t]
\centering
\includegraphics[width=\columnwidth]{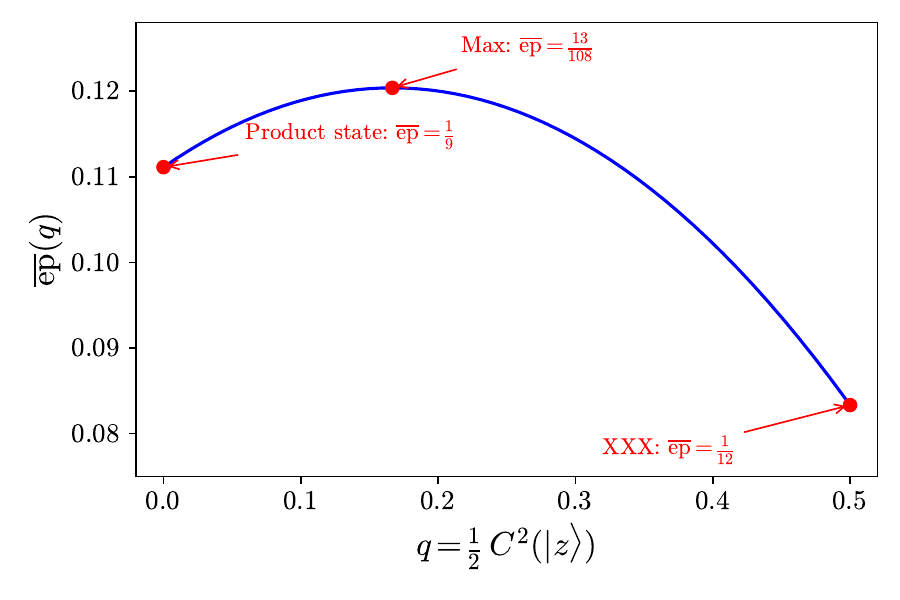}
\caption{{\em Time-averaged entangling power of the
spin-1/2 isospectral family,
Eq.~(\ref{eq:ep_isospectral}), as a function of
$q = C^2(|z\>)/2$, where $C$ is the concurrence of
the projector state $|z\>$.
The product-state limit ($q = 0$, $\avg{\ep} = 1/9$)
and the XXX/Bell-state limit
($q = 1/2$, $\avg{\ep} = 1/12$) are marked, together
with the maximum at $q = 1/6$.
The XXX model sits at the minimum of
$\avg{\ep}$ over the isospectral family.
}}
\label{fig:ep_isospectral}
\end{figure}

\paragraph{Analytic formula for the isospectral family.}
The local-unitary structure above already determines on which
property of $|z\>$ the entangling power can depend.
Under a local unitary $|z\> \mapsto (u_A \otimes u_B)\,|z\>$,
the Hamiltonian in Eq.~(\ref{eq:H_iso}) transforms as
$H \mapsto (u_A \otimes u_B)\,H\,(u_A \otimes u_B)^\dagger$,
and since $\avg{\ep}$ is invariant under such conjugation,
$\avg{\ep}[H(|z\>)]$ is a local-unitary invariant of $|z\>$.
For two-qubit pure states, every entanglement measure is a
local-unitary invariant, and all such invariants are
functionally equivalent to the concurrence $C(|z\>)$~\cite{Hill:1997pfa,Wootters:1997id}, as shown in Ref.~\cite{Low:2021ufv}. Thus $\avg{\ep}$ depends
on $|z\>$ only through $C(|z\>)$, and we parametrize the
family by
\be
\label{eq:q_def}
q 
= \tfrac{1}{2}\,C^2(|z\>)\,.
\ee
The parameter $q$ ranges from $0$ (product state,
$\alpha = 1$) to $1/2$ (Bell state,
$\alpha = \beta = 1/\sqrt{2}$). Following the algorithm in Sec.~\ref{subsect:time}, we obtain
the time-averaged entangling power
\be
\label{eq:ep_isospectral}
\avg{\ep}(q) = \frac{1 + q - 3q^2}{9}\,.
\ee
The product-state limit $q = 0$ gives
$\avg{\ep} = 1/9$, the Bell-state
(XXX) limit $q = 1/2$ gives
$\avg{\ep} = 1/12$, and the maximum
$\avg{\ep} = 13/108$ occurs at $q = 1/6$.

The non-monotonic dependence on $q$ is noteworthy: the
entangling power initially {\em increases} as $|z\>$
acquires entanglement, reaches a maximum at $q = 1/6$,
and then decreases as $|z\>$ approaches the Bell state.
The XXX model,
with maximally entangled eigenstates, sits at the
{\em minimum} of $\avg{\ep}$ over the isospectral
family, confirming that the suppression of entanglement
power at the $SU(2)$ point is driven by eigenvector
structure, not just eigenvalue degeneracy.
This non-monotonic behavior is displayed in
Fig.~\ref{fig:ep_isospectral}.

The suppression of entangling power at the non-Abelian
$SU(2)$ point complements Ref.~\cite{Majidy:2022kzx}, which
found that noncommuting charges can instead enhance the
entanglement of typical constrained states. Together, these
results suggest that non-Abelian symmetry can affect operator-
and state-level entanglement in qualitatively different ways.

\begin{figure*}[t]
\centering
\includegraphics[width=\textwidth]{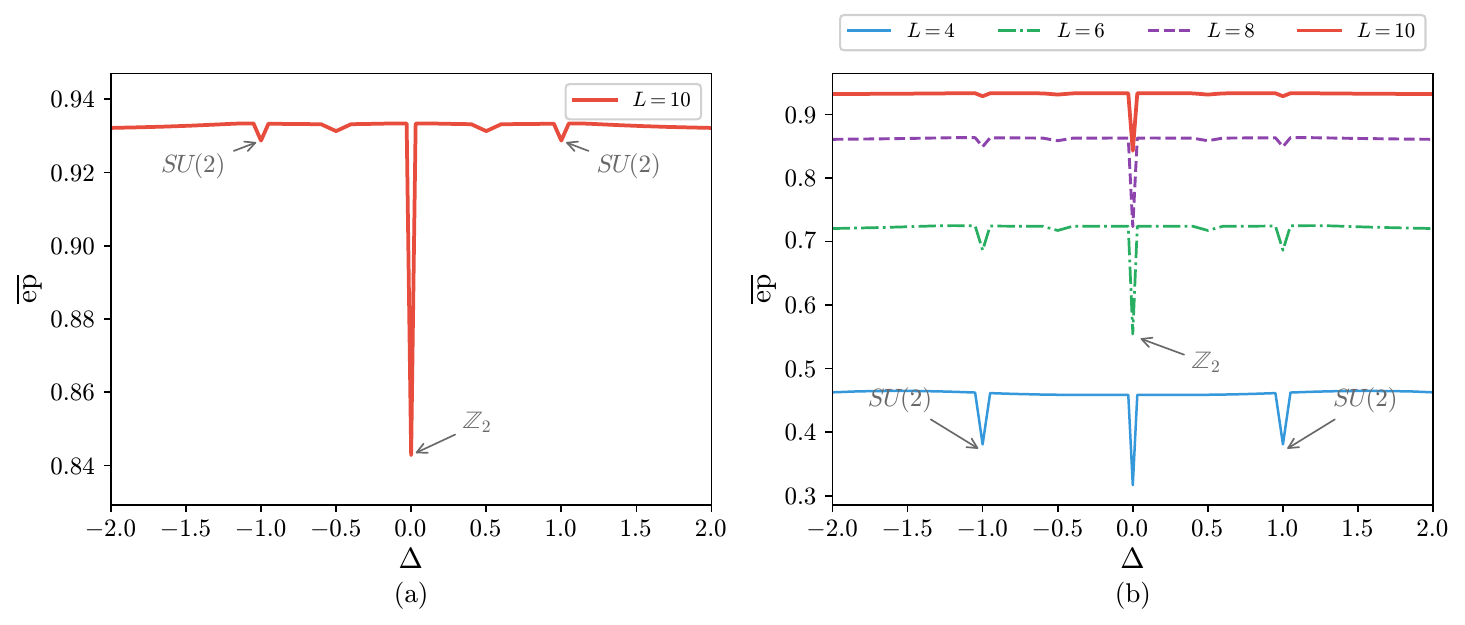}
\caption{\em Time-averaged entangling power $\avg{\ep}$
for the spin-$1/2$ XXZ chain as a function of anisotropy
$\Delta$.
(a) $L = 10$ chain, shown on a scale to resolve the residual dip
structure near saturation.
(b)~Overlay of $L = 4$ (solid blue), 6 (dash-dotted
green), 8 (dashed purple), and 10 (solid red).
The $L = 2$ case was already shown in Fig.~\ref{fig:ep_xxz_analytic}.
Root-of-unity fine structure at $\Delta = \pm 1/2$
becomes visible for $L \geq 6$.
\label{fig:spin12_overlay}}
\end{figure*}

\section{Beyond Two-site Models}
\label{sec:finitesize}

The two-site results of the preceding section establish a clear
correlation between symmetry enhancement and entanglement
suppression. We now study this correlation  beyond two-site models. In this Section we
compute $\avg{\ep}$ for spin chains of increasing length $L$ and study the interplay between
 the growth of the Hilbert space and the richness of the
many-body spectrum.

We consider spin-$1/2$ and spin-$1$ XXZ models~\cite{Orbach:1958} with open boundary conditions:
\be
\label{eq:H_chain}
H_{\rm XXZ} = \sum_{j=1}^{L-1}\bigl(
S_x^{(j)}S_x^{(j+1)} + S_y^{(j)}S_y^{(j+1)}
+ \Delta\, S_z^{(j)}S_z^{(j+1)}\bigr)\,,
\ee
where $S_i^{(j)}$ are the spin-$s$ angular momentum
operators at site $j$, with $s = 1/2$ or $s = 1$.
We choose the equal bipartition
$\{1,\ldots,L/2\}|\{L/2+1,\ldots,L\}$,
so that $d_A = d_B = (2s+1)^{L/2}$.
The full Hilbert space has dimension $d = (2s+1)^L$:
for spin-$1/2$, the largest system we study is $L = 10$ ($d = 1024$),
and for spin-$1$, $L = 6$ ($d = 729$).
Time-averaged entangling power is computed by
numerically diagonalizing $H$ and then performing
the infinite-time average exactly using the
eigenvalue-grouping algorithm described in
Sec.~\ref{subsect:time}.

\subsection{Spin-$1/2$ chains}
\label{sec:spin12_chain}

Figure~\ref{fig:spin12_overlay} shows $\avg{\ep}$
as a function of $\Delta$ for spin-$1/2$ XXZ chains
at $L = 4$, 6, 8, and 10.
Three dips at $\Delta = -1$, 0, and 1 are clearly
visible for all chain lengths, confirming that the
symmetry--entanglement correlation. The most striking feature is the dip at the XX point $\Delta=0$, which is not present at $L=2$ in Fig.~\ref{fig:ep_xxz_analytic}. (This is due to an accidental cancellation at $\Delta=0$ for the two-site model.) In addition, the depth of the dips at the XXX point decreases rapidly as $L$ increases, while the XX dip remains quite pronounced even at $L=10$.

The rapid decay of the $SU(2)$ dip with system size
has a transparent physical origin.
For $L = 2$ the XXX model has only two
$SU(2)$ multiplets ($S = 0$ and $S = 1$), and the
eigenstates are entirely determined by Clebsch--Gordan
coefficients. The symmetry  constrains both the
eigenvalue degeneracy pattern and the eigenvector
structure, producing a large suppression of $\avg{\ep}$.

As $L$ grows, however, the total-spin decomposition
$(2s+1)^{\otimes L} = \bigoplus_{S} m_S \cdot (2S+1)$
produces an increasing number of multiplets whose
multiplicities $m_S$ grow rapidly.
For $L$ spin-$1/2$ particles, the allowed
total-spin values range from $S = 0$ (or $1/2$)
to $S = L/2$, so the number of distinct sectors
grows only linearly, as $L/2 + 1$.
The multiplicities, on the other hand, must
account for the full $2^L$-dimensional Hilbert space:
$\sum_S m_S (2S+1) = 2^L$.
Since there are only $O(L)$ sectors and the
largest dimension $(2S+1)$ is at most $L + 1$,
the dominant multiplicities must grow exponentially
to fill the Hilbert space.
Indeed, for the singlet sector
$m_0 = \binom{L}{L/2} - \binom{L}{L/2 + 1}
\sim 2^L / L^{3/2}$ by Stirling's approximation.

The $SU(2)$ symmetry labels each eigenstate by its
global quantum numbers $(S, M)$ but does not constrain
the Hamiltonian dynamics {\em within} each multiplicity
space of dimension $m_S$.
The fraction of the Hilbert space controlled by
the symmetry labels is therefore of order
$L/2^L$, which vanishes exponentially.
Concretely, at $L = 2$ the singlet sector has
$m_0 = 1$---the symmetry completely fixes the
eigenstate---while at $L = 10$, $m_0 = 42$, and
the Hamiltonian is free to choose among 42
linearly independent singlet states.
It is this unconstrained dynamics within the
multiplicity spaces that washes out the
entanglement suppression.

It is worth emphasizing that the dip does not
disappear because the symmetry itself is
weakened---$SU(2)$ holds at $\Delta = \pm 1$ for
all $L$---but because the {\em relative} importance
of the symmetry constraint diminishes as the
Hilbert space grows exponentially.
Numerically, this is demonstrated in
Fig.~\ref{fig:spin12_overlay}~(a) for $L=10$.

\subsection{XX dip at $\Delta=0$}
\label{sec:free_fermion}

In striking contrast to the rapid decay of the
XXX dip, the XX dip persists to significantly
longer chains.
The mechanism is qualitatively different and can be
traced to the free-fermion structure at
$\Delta = 0$.

The Jordan--Wigner transformation~\cite{Jordan:1928wi,LiebSchultzMattis:1961}
maps spin-$1/2$ operators to spinless fermion
operators $c_j$, $c_j^\dagger$ via
\bea
\label{eq:JW}
S_j^+ &=& c_j^\dagger \prod_{l<j}(1-2n_l)\,,
\quad
S_j^- = \bigl(S_j^+\bigr)^\dagger\,,
\non\\
S_j^z &=& n_j - \tfrac{1}{2}\,,
\eea
where $n_j = c_j^\dagger c_j$ is the fermion number
operator.
For nearest-neighbor sites the Jordan--Wigner string
cancels, and the XXZ
Hamiltonian~(\ref{eq:H_chain}) becomes~\cite{Lieb:1961fr}
\be
\label{eq:H_JW}
H = \sum_{j=1}^{L-1}\left[
\tfrac{1}{2}\bigl(c_j^\dagger c_{j+1} + {\rm h.c.}\bigr)
+ \Delta\bigl(n_j - \tfrac{1}{2}\bigr)
\bigl(n_{j+1} - \tfrac{1}{2}\bigr)\right].
\ee
At $\Delta = 0$ the quartic interaction term vanishes
and the Hamiltonian reduces to free fermions hopping
on a chain of $L$ sites with open boundary conditions.
The resulting single-particle energies are
$\varepsilon_k = \cos(k\pi/(L+1))$ for
$k = 1, 2, \ldots, L$, and each many-body energy
eigenvalue is a sum of a subset of these
single-particle levels:
$E = \sum_{k \in \mathcal{S}} \varepsilon_k$,
where $\mathcal{S}$ labels the occupied
single-particle states.

The free-fermion structure has a direct consequence
for the time-averaged entangling power.
The time-averaging algorithm of
Sec.~\ref{subsect:time} groups eigenstate pairs
by their energy differences:
pairs with the same $E_\alpha - E_\beta$
contribute coherently to the purity sums.
For a generic interacting Hamiltonian with no special
structure, the $d^2 = 4^L$ energy differences
$E_\alpha - E_\beta$ are generically all distinct,
so nearly every group contains a single pair.
Single-pair groups contribute independently to the
purity sums---there are no large coherent
contributions that could enhance the purity---and
the time-averaged entangling power remains close
to the Haar-random value.
At the free-fermion point the situation is
qualitatively different.
Since every many-body energy is a sum of
single-particle levels $\varepsilon_k$, every energy
difference $E_\alpha - E_\beta$ is itself a sum or
difference of at most $L$ values drawn from a set of
size~$L$.
The number of distinct energy differences therefore
grows only polynomially, as $O(L^3)$, even though the
number of eigenstate pairs grows exponentially as
$O(4^L)$.
Exponentially many pairs are thus forced into
polynomially many groups, creating large degenerate
groups whose coherent contributions {\em can}
enhance the purity sums.
Whether and by how much these contributions actually
suppress $\avg{\ep}$ depends on the
Hilbert--Schmidt overlaps of the $M$-matrices
within each group---a quantity not determined
by the counting argument alone---but the numerical
data confirm that the suppression is substantial.
Because the polynomial-versus-exponential
mismatch persists for all $L$, the XX dip
decays much more slowly than the XXX dip.

The free-fermion structure also explains a second,
more subtle feature of the XX dip.
The single-particle energies
$\varepsilon_k = \cos(k\pi/(L+1))$ satisfy
$\varepsilon_k + \varepsilon_{L+1-k} = 0$,
so the single-particle spectrum is symmetric about
zero.
Particle--hole conjugation therefore maps every
many-body eigenvalue $E$ to $-E$, endowing the
Hamiltonian with a $\mathbb{Z}_2$
spectrum-flipping symmetry.
In the spin language this symmetry is generated by
$W = e^{i\pi S_{z,1}} \otimes I \otimes
e^{i\pi S_{z,3}} \otimes I$, the tensor product of
the single-site $\pi$-rotation of
Sec.~\ref{sec:spin_one} applied to alternating
(odd-numbered) sites.
The operator $W$ acts on one site
from each subsystem, and
the $E \leftrightarrow -E$ pairing forces additional
spectral degeneracies beyond those already required
by the free-fermion group counting, further enhancing
the purity sums.

 It is illuminating to decompose
the entangling power into its two purity sums,
$\avg{I_0}$ and $\avg{I_1}$, associated with the
transposition operators $T_{13}$ (the $A$-swap) and
$T_{24}$ (the $B$-swap) in the doubled Hilbert space,
respectively.
For a symmetric bipartition ($d_A = d_B$), one might
expect these two contributions to be comparable.
At the XXX point ($\Delta = 1$) this is
indeed the case: $SU(2)$ symmetry treats $T_{13}$ and
$T_{24}$ on equal footing, producing
$\delta\avg{I_0} = \delta\avg{I_1}$ relative to the
generic XXZ values.
The $SU(2)$ dip is therefore a {\em symmetric}
enhancement of both purity contributions.
At the XX point, however, the particle--hole generator
$W$ couples differently to $T_{13}$ than to $T_{24}$:
it acts on sites 1 and 3
(one from $A$ and one from $B$), breaking the
$A \leftrightarrow B$ symmetry.
The result is a pronounced asymmetry
$\delta\avg{I_0} \gg \delta\avg{I_1}$---for $L = 4$,
 $\delta\avg{I_0}/\delta\avg{I_1}$ is
approximately tenfold---driving the deep XX dip.

This asymmetric mechanism is invisible at $L = 2$,
where the universal Bell-state eigenvectors cause
$\delta\avg{I_0}$ and $\delta\avg{I_1}$ to cancel
exactly, producing no net dip.
For $L \geq 4$, the eigenvectors become
coupling-dependent, the cancellation is broken, and the
particle--hole symmetry generates a large, asymmetric
enhancement of $\avg{I_0}$ that grows with $L$ relative
to the XXX contribution.

We note that, at $\Delta=0$,  the spin-$1/2$ XX point also enjoys a
non-invertible lattice symmetry implementing T-duality
of the compact-boson IR theory~\cite{Pace:2024tduality}.\footnote{We thank Shu-Heng Shao for bringing
Ref.~\cite{Pace:2024tduality} to our attention.}
This non-invertible structure is related to the
free-fermion integrability exploited above and is
special to the Jordan--Wigner algebra of spin-$1/2$
chains.

\subsection{Quantum Group at Root-of-unity}
\label{sec:root_of_unity}

In addition to the striking dips at $\Delta = 0$ and
$\Delta = \pm 1$, fine structure appears at
$\Delta = \pm 1/2$ for $L \geq 6$, visible in
Fig.~\ref{fig:spin12_overlay}.
These additional features have a purely algebraic origin:
they arise from a hidden symmetry that the XXZ model
possesses for {\em all} values of $\Delta$, not
only at the XXX point.

At the isotropic point $\Delta = 1$, the XXZ Hamiltonian
has $SU(2)$ symmetry and the Hilbert space decomposes
into familiar spin-$j$ multiplets.
Away from the isotropic point, the full $SU(2)$ is
broken, but a deformed version survives.
The spin-$1/2$ XXZ Hamiltonian with open boundary
conditions commutes with the generators of
$U_q(\mathfrak{sl}_2)$, the quantum group
introduced by Drinfel'd~\cite{Drinfeld:1985} and
Jimbo~\cite{Jimbo:1985zk} as a
one-parameter deformation of
$\mathfrak{sl}_2$, the Lie algebra of
$SU(2)$.
(See, e.g., Ref.~\cite{Chari:1994pz, Lamers:2015qig}
for pedagogical introductions.)
The deformation parameter $q$ is related to $\Delta$  by
\begin{equation}
\Delta = \cos\gamma\,, \qquad q = e^{i\gamma}\,,
\label{eq:q_root}
\end{equation}
so that $q = 1$ at the XXX point
$\Delta = 1$ and $q = i$ at the free-fermion point
$\Delta = 0$.
Concretely, the deformed algebra is generated by
raising and lowering operators $J_\pm$ together with
an exponentiated Cartan generator $K = q^{J_z}$,
satisfying the $q$-deformed commutation relations
\begin{equation}
K J_\pm K^{-1} = q^{\pm 1} J_\pm\,, \qquad
[J_+, J_-] = \frac{K^2 - K^{-2}}{q - q^{-1}}\,.
\label{eq:q_comm}
\end{equation}
In the limit $q \to 1$ the right-hand side of the
second relation reduces to $2J_z$, recovering the
standard $\mathfrak{sl}_2$ algebra.
The key point is that, for generic values of $q$,
the representation theory of
$U_q(\mathfrak{sl}_2)$ mirrors that of
$\mathfrak{sl}_2$: irreducible representations are
labeled by non-negative half-integers $j$, each of
dimension $2j + 1$, and the spectrum of the XXZ
chain decomposes into $U_q(\mathfrak{sl}_2)$
multiplets in close analogy with the $SU(2)$ case.
In this sense the XXZ model retains a
``deformed $SU(2)$'' symmetry for all $\Delta$,
which is the algebraic reason underlying its
integrability.

The situation changes drastically when $q$ is a
root of unity, i.e., when $q^N = 1$ for some
positive integer
$N$~\cite{Pasquier:1989kd,Alcaraz:1987zr,Deguchi:2000nn}.
In ordinary $SU(2)$ representation theory, every
representation is either irreducible or decomposes
as a direct sum of irreducible representations.
At roots of unity, this familiar structure breaks
down: the center of the algebra enlarges,
and representations of spin $j \geq N/2$ become
{\em reducible but indecomposable}---they contain
invariant subspaces but cannot be split into a
direct sum of irreducibles.
To illustrate: in ordinary $SU(2)$, if a
representation contains a spin-$j$ subspace, one
can always find an orthogonal complement that is
itself invariant, decomposing the full space as
$V_j \oplus V_{j'}$.
At roots of unity this complementary subspace does
not exist---the action of the generators on vectors
outside the invariant subspace necessarily mixes
them back in.
As a result, multiplets that would form independent
irreducible representations at generic $q$ are
forced into a single indecomposable block at roots
of unity.
Since the Hamiltonian commutes with the quantum
group generators, it must act as a scalar on each
indecomposable block; the previously distinct
multiplets are therefore constrained to share
the same energy eigenvalue, producing degeneracies
that have no counterpart in the ordinary
$\mathfrak{sl}_2$ theory.

\begin{figure}[t]
\centering
\includegraphics[width=\columnwidth]{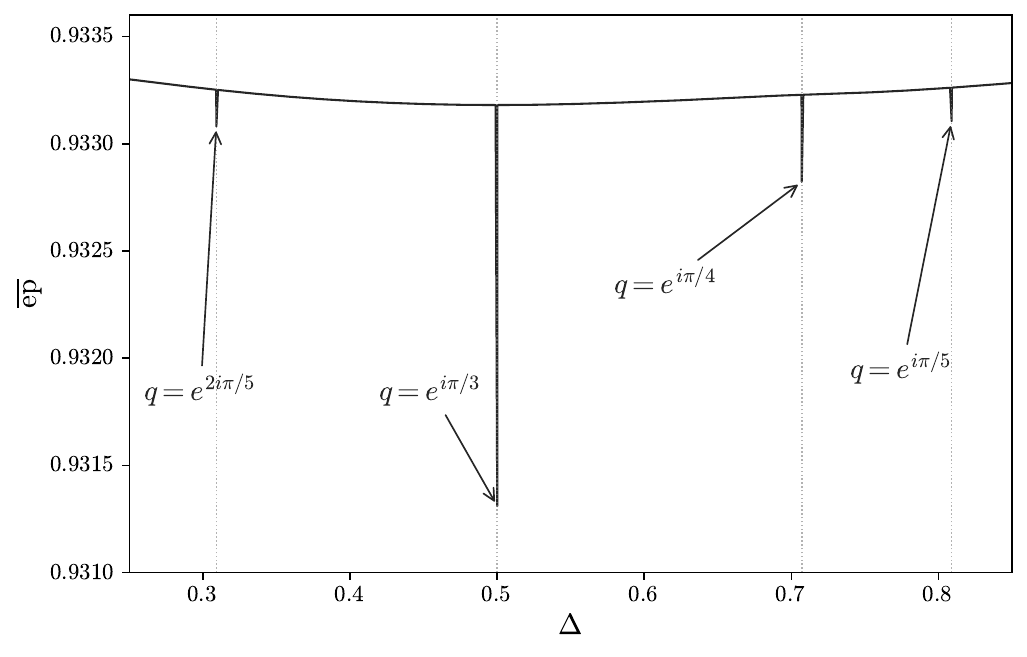}
\caption{\em Dense scan of $\avg{\ep}$ for the
spin-$1/2$ XXZ chain at $L = 10$ over
$\Delta \in [0.25, 0.85]$, resolving
all four root-of-unity dips.
Vertical dashed lines mark the exact algebraic
locations.
\label{fig:fine_scan}}
\end{figure}

To be concrete, consider $\Delta = 1/2$, which
corresponds to $\gamma = \pi/3$ and
$q = e^{i\pi/3}$, a primitive sixth root of unity
($q^6 = 1$).
For a chain of length $L \geq 6$, the Hilbert
space is large enough to accommodate representations
with $j \geq 3$, and these higher-spin
representations become indecomposable.
The resulting degeneracies are additional to those
already present from $U_q(\mathfrak{sl}_2)$ at
generic $q$: eigenvalues that would be distinct at
nearby values of $\Delta$ become exactly degenerate
at $\Delta = 1/2$.
Within the time-averaging framework of
Sec.~\ref{subsect:time}, these additional degeneracies
enlarge the $\omega$-groups and thereby create the
{\em potential} for further suppression of~$\avg{\ep}$;
the numerical results below (see Fig.~\ref{fig:fine_scan})
confirm that this potential is realized.

This algebraic picture makes three testable
predictions.
First, the width of the dip should be extremely
narrow, reflecting the fact that the indecomposable
structure exists only exactly at the root-of-unity
point; any deviation from $q^N = 1$ restores the
generic representation theory and lifts the
additional degeneracies.
Second, the dips should appear symmetrically at
$\Delta = \pm 1/2$, because the transformation
$\gamma \to \pi - \gamma$ maps $\Delta \to -\Delta$
while preserving the root-of-unity condition.
Third, the dips should become visible only for
$L \geq 6$, since shorter chains do not have
representations of sufficiently high spin to trigger
the indecomposable structure.
A complete analytical prediction for the
root-of-unity dip depth would require computing
the Hilbert--Schmidt overlaps of $M$-matrices within
each indecomposable block---a nontrivial
representation-theoretic calculation that we leave
for future work.

To confirm the root-of-unity prediction at higher $N$
and longer chains, we perform a dense scan of 
$\Delta$ values in the interval $[0.25, 0.85]$,
with the exact algebraic values
$\Delta = 1/2$ ($N = 3$, 6th root),
$\Delta = 1/\sqrt{2}$ ($N = 4$, 8th root),
$\Delta = \cos(2\pi/5)$ ($N = 5$, 10th root), and
$\Delta = \cos(\pi/5)$ ($N = 5$, 10th root)
explicitly included in the grid.
Fig.~\ref{fig:fine_scan} shows the results for
$L = 10$.
All four special points exhibit clear dips centered
precisely at the predicted algebraic values.
The dip depths follow the expected $N$-hierarchy:
the $N = 3$ dip is the deepest, followed by
$N = 4$, while the two $N = 5$ windows
produce the shallowest dips.
At $L = 8$ the same hierarchy holds with
 the dips becoming
shallower for increasing $L$.

This decrease of the root-of-unity dip depths
with $L$ parallels the behavior of the XXX dip
rather than the free-fermion dip: both are driven by
symmetry constraints that become subdominant as the
Hilbert space dimension grows.
The hierarchy in $N$ can be understood from the observation that higher-order roots
of unity require representations of higher spin before
the indecomposable structure becomes operative, so the
corresponding dips need increasingly long chains to
develop.

\begin{figure*}[t]
\centering
\includegraphics[width=\textwidth]{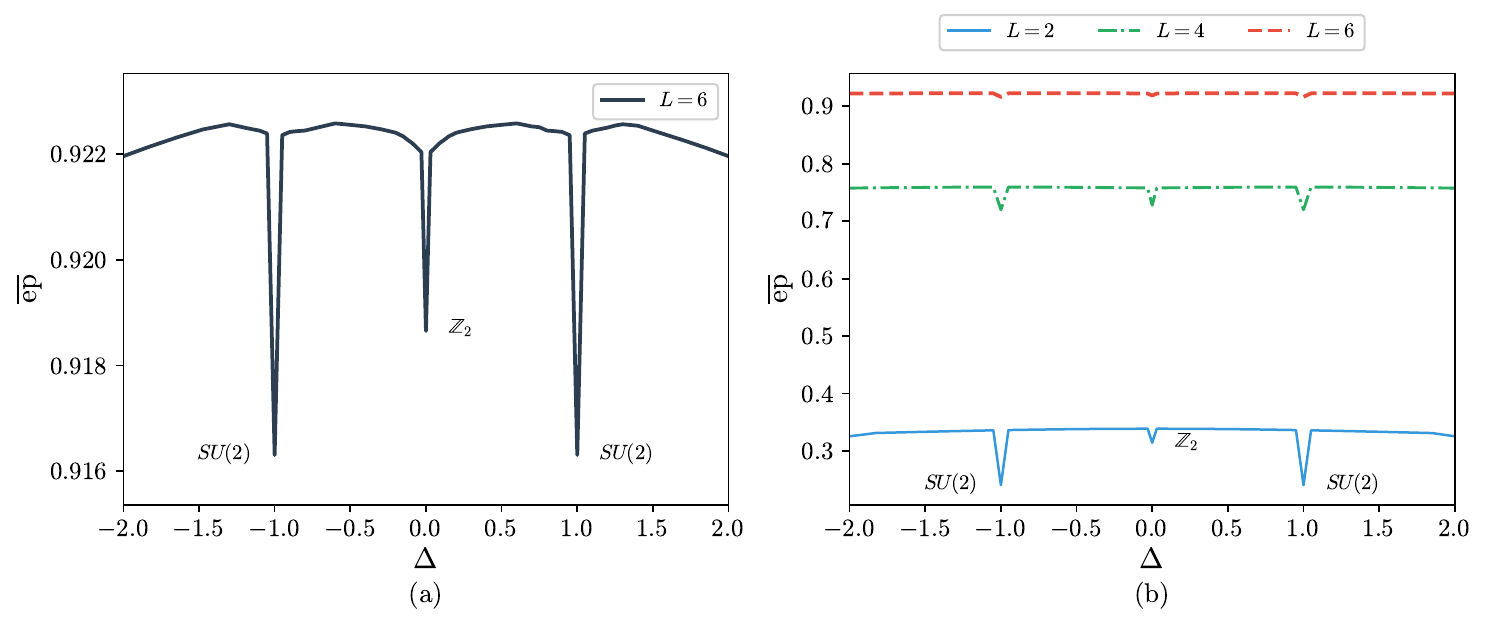}
\caption{\em Time-averaged entangling power $\avg{\ep}$
for the spin-$1$ XXZ chain as a function of anisotropy
$\Delta$.
(a)~Zoom in for $L = 6$.
(b)~Overlay of $L = 2$ (solid blue), 4 (dash-dotted
green), and 6 (dashed red).
All three dips at $\Delta = -1$, $0$, and $1$ are
visible but decay rapidly with $L$.
\label{fig:spin1_overlay}}
\end{figure*}

\subsection{Spin-$1$ chains}
\label{sec:spin1_chain}

We extend the calculation to spin-$1$ XXZ chains
with $L = 2$, 4, and 6 sites.
Fig.~\ref{fig:spin1_overlay} shows $\avg{\ep}$
as a function of $\Delta$ for the three chain lengths.
The qualitative picture is the same as spin-$1/2$:
three dips at $\Delta = -1$, 0, and 1 persist for all
chain lengths studied, and the dip depths decrease as
$L$ increases.

Two features distinguish the spin-$1$ results from
spin-$1/2$.
First, the XX dip at $\Delta=0$ is already visible at $L=2$, contrary to the $L=2$, spin-1/2 case.  This difference traces back to the two-site
eigenvector structure: for spin-$1/2$, the Bell-state
eigenvectors are universal and coupling-independent,
producing the exact $I_0/I_1$ cancellation that renders
the XX dip invisible at $L = 2$.
For spin-$1$, the eigenvectors are already
coupling-dependent at $L = 2$, so the $\mathbb{Z}_2$
symmetry at $\Delta = 0$ produces a visible 7\% dip
even in the minimal model.

Second, and perhaps most strikingly, the {\em relative}
depth of the XX and XXX dips evolves in the
opposite direction compared with spin-$1/2$.
For spin-$1/2$ chains, the free-fermion mechanism at
$\Delta = 0$ causes the XX dip to persist and even
deepen relative to the XXX dip as $L$ grows:
by $L = 10$ the XX dip  is more than twice the
XXX dip.
For spin-$1$, the situation is reversed: the XX dip
at $\Delta = 0$ decays {\em faster} than the XXX
dip at $\Delta = \pm 1$.
The contrast with spin-$1/2$ can be traced to the
absence of a simple free-fermion mapping for the
spin-$1$ chain: the Jordan--Wigner transformation
applies only to spin-$1/2$ systems, so the
polynomial-versus-exponential counting argument of
Sec.~\ref{sec:free_fermion} does not carry over.
As a result, the XX dip in the spin-$1$ chain
decays at a rate comparable to the XXX dip.
More generally, all dips decay faster with $L$ for
spin-$1$ than for spin-$1/2$.
This is expected on general grounds: the larger
local Hilbert space ($d_{\rm loc} = 3$ versus 2) means
$d = 3^L$ grows faster than $2^L$, and the symmetry
constraints become subdominant more rapidly.

\section{Entangling Power and Integrability}
\label{sec:integrability}

The XXZ chain is Bethe-ansatz integrable for all values
of $\Delta$, so the symmetry dips documented in
Sec.~\ref{sec:finitesize} are features that appear
{\em within} an integrable family.
A natural question is whether the entangling power
can also detect integrability, or whether the dips it
reveals are purely symmetry-driven.
We address this question from two complementary
directions.
First, we break integrability while preserving the
symmetry structure by adding a next-nearest-neighbor
coupling (the $J_1$-$J_2$ model).
Second, we explore a different integrable family---the
bilinear-biquadratic spin-$1$ chain---that contains
an $SU(3)$-symmetric point and exhibits a qualitatively
different degeneracy structure.

\subsection{Integrability breaking via the $J_1$-$J_2$
coupling}
\label{sec:j1j2}

Consider the spin-$1/2$ $J_1$-$J_2$ XXZ chain with
open boundary conditions:
\be
\label{eq:H_j1j2}
H = J_1\ H_{\rm XXZ} 
+ J_2 \sum_{j=1}^{L-2}
\vec{S}^{(j)} \cdot \vec{S}^{(j+2)}\, ,
\ee
where $H_{\rm XXZ}$ is the Hamiltonian in Eq.~(\ref{eq:H_chain}).
At $J_2 = 0$ the model reduces to the integrable XXZ
chain.
For $J_2 \neq 0$, integrability is generically broken.
We set $J_1 = 1$ throughout.

\begin{figure*}[t]
\centering
\includegraphics[width=\textwidth]{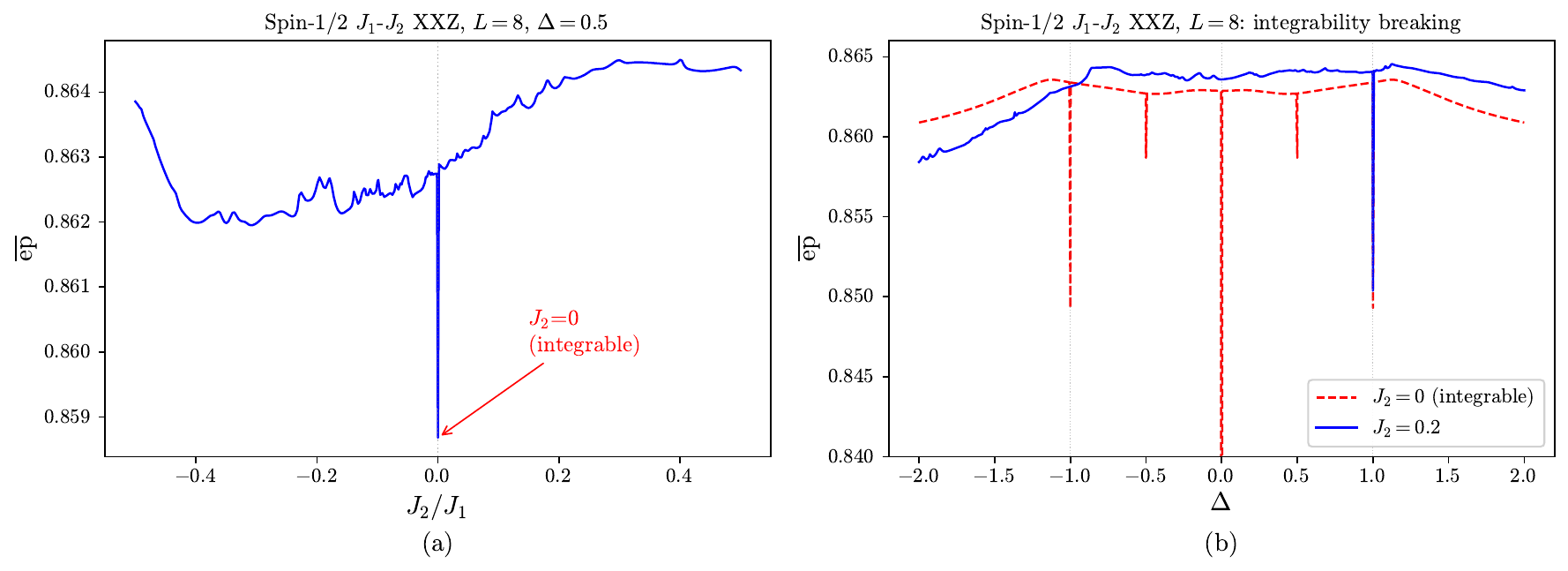}
\caption{\em Time-averaged entangling power for the
spin-$1/2$ $J_1$-$J_2$ XXZ chain at $L = 8$.
(a)~$\avg{\ep}$ as a function of the integrability-breaking
parameter $J_2/J_1$ at fixed $\Delta = 0.5$.
A dip at $J_2 = 0$ (the integrable point) is visible,
with $\avg{\ep}$ approximately $0.5\%$ below the
generic non-integrable plateau.
(b)~$\avg{\ep}$ as a function of anisotropy $\Delta$,
comparing $J_2 = 0$ (integrable, dashed red)
and $J_2 = 0.2$ (non-integrable, solid blue).
The symmetry dips at $\Delta = 0$ and $\Delta = -1$ are
almost entirely eliminated by the integrability-breaking
perturbation, while the XXX dip at $\Delta = 1$
survives.
\label{fig:j1j2}}
\end{figure*}

Figure~\ref{fig:j1j2}(a) shows $\avg{\ep}$ as a function
of $J_2$ at fixed $\Delta = 0.5$ for $L = 8$.
A clear dip is visible at $J_2 = 0$: the integrable
point has $\avg{\ep} \approx 0.859$, approximately
$0.5\%$ below the generic non-integrable values
($\avg{\ep} \approx 0.863$).
Again the dip is very sharp:
dense numerical scans with 300 uniformly-spaced
$J_2$ values in $[-0.5,\, 0.5]$ reveal that at
$|J_2| \gtrsim 0.002$ the entangling power has already
returned to its generic non-integrable value,
and only the \emph{exactly} integrable point $J_2 = 0$
shows any suppression.

Figure~\ref{fig:j1j2}(b) compares the $\Delta$-dependence
of $\avg{\ep}$ at $J_2 = 0$ and $J_2 = 0.2$ for $L = 8$.
The results reveal a clear hierarchy of robustness among
the three dip features:
both the XX dip at $\Delta = 0$ and 
the $\Delta = -1$ dip  nearly disappeared for $J_2 = 0.2$, while
the XXX dip at $\Delta = 1$ survives and even
deepens slightly.
The survival of the XXX dip is expected:
the isotropic $J_1$-$J_2$ XXX model at
$\Delta = 1$ retains full $SU(2)$ symmetry regardless
of $J_2$, while the surrounding XXZ points still have
only $U(1)$.
In contrast, the $\mathbb{Z}_2$ spectrum-flipping symmetry
$W = e^{i\pi S_{z,1}} \otimes I \otimes e^{i\pi S_{z,3}}
\otimes I$ that generates the XX dip acts on
nearest-neighbor pairs and is destroyed by the $J_2$
coupling.

\subsection{Bilinear-biquadratic spin-$1$ chain}
\label{sec:blbq}

In this subsection we study
the bilinear-biquadratic spin-$1$ chain, which  provides another 
controlled setting to study the interplay between entangling power and integrability.

The most general isotropic nearest-neighbor spin-$1$
Hamiltonian (up to an overall scale) can be
parametrized~as
\be
\label{eq:H_blbq}
H = \sum_{j=1}^{L-1} \biggl[
\cos\theta\;\vec{S}_j \cdot \vec{S}_{j+1} 
+ \sin\theta\;
\bigl(\vec{S}_j \cdot \vec{S}_{j+1}\bigr)^{\!2}
\biggr]\,,
\ee

\begin{figure*}[t]
\centering
\includegraphics[width=\textwidth]{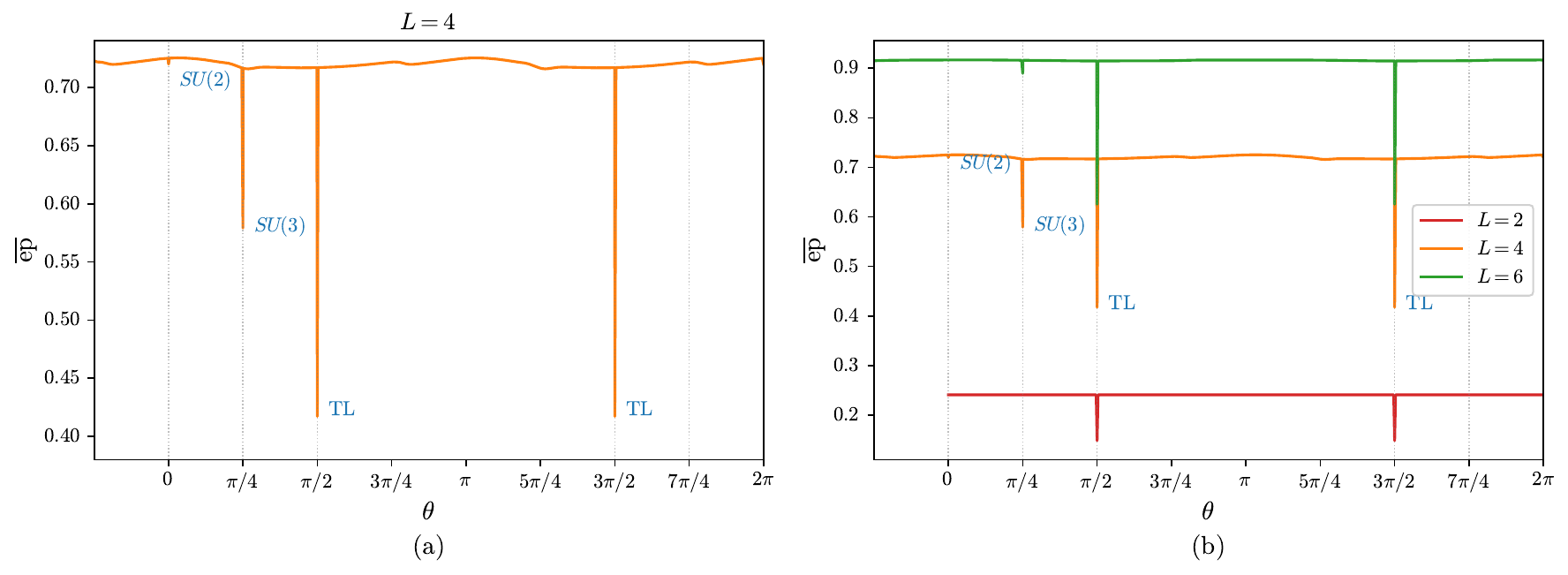}
\caption{\em Time-averaged entangling power $\avg{\ep}$
for the bilinear-biquadratic spin-$1$
chain~(\ref{eq:H_blbq}) as a function of $\theta$.
(a)~$L = 4$, where the full dip hierarchy is most
clearly resolved: Temperley--Lieb (TL, $\theta = \pi/2, 3\pi/2$), $SU(3)$ ($\theta = \pi/4$),
and $SU(2)$ ($\theta = 0$).
(b)~Overlay of $L = 2$ (red), $L = 4$ (orange),
and $L = 6$ (green).
\label{fig:blbq}}
\end{figure*}

\noindent
where $\theta \in [0, 2\pi)$ parametrizes the ratio of
the bilinear and biquadratic couplings.
This model has a rich phase
diagram~\cite{Haldane:1983ru,Affleck:1987vf,Affleck:1988xj}
with several special integrable points:
$\theta = 0$ is the isotropic XXX chain, with
$SU(2)$ symmetry;
$\theta = \pi/4$ is the Uimin--Lai--Sutherland (ULS)
point~\cite{Uimin:1970,Lai:1974,Sutherland:1975},
at which the bilinear and biquadratic couplings are equal
and the symmetry is enhanced from $SU(2)$ to $SU(3)$;
$\theta = -\pi/4$ (equivalently $7\pi/4$) is the
Takhtajan--Babujian
point~\cite{Takhtajan:1982,Babujian:1982}, which is also
integrable;
and $\theta = \pi/2$ is the pure biquadratic chain.

Figure~\ref{fig:blbq} shows $\avg{\ep}$ as a function
of $\theta$ for chains of length $L = 2$, 4, and~6.
Away from the special points, $\avg{\ep}$ forms a high
plateau, punctuated
by sharp dips at the biquadratic,
ULS, and XXX points.
The most dramatic feature is the deep dip at the pure
biquadratic point $\theta = \pi/2$,
at which point the Hamiltonian is purely biquadratic,
$H = \sum_j (\vec{S}_j \cdot \vec{S}_{j+1})^2$,
and for $L = 2$ the spectrum has the degeneracy
pattern $\{1, 1, 1, 1, 1, 1, 1, 1, 4\}$---whose
algebraic origin in the Temperley--Lieb algebra we
explain below---which is far more
degenerate than the XXX pattern $\{1, 3, 5\}$.
This extreme degeneracy suppresses the entangling power
by eliminating nearly all oscillatory terms in the
time average. In contrast to the biquadratic dip, the ULS point
at $\theta = \pi/4$ shows a moderate but clearly visible
suppression, while the XXX point at $\theta = 0$ shows much smaller
dips.

It is worth pointing out that the Takhtajan--Babujian
point at $\theta = -\pi/4$ (equivalently $7\pi/4$) shows
no discernible dip at any chain length studied: $\avg{\ep}$
at this point is indistinguishable from the generic
plateau to within $0.2\%$.
This is a striking result given that the TB point is
integrable and solvable by Bethe ansatz, just as the
XXX chain.
However, the spectrum at $\theta = 7\pi/4$ does not
exhibit the enhanced degeneracies that suppress the
time-averaged entangling power at the other special
points. It would be interesting to study if an entanglement probe exists for the TB point.

Given the strong biquadratic
suppression due to the extreme eigenvalue degeneracy
at $\theta = \pi/2$, it is worth explaining
its algebraic origin.
For two spin-$1$ particles on neighboring sites
$i$ and $i\!+\!1$, the total spin
$S_{\rm tot}$ can be 0, 1, or~2, and  the biquadratic
interaction can be written as
$(\vec{S}_i \cdot \vec{S}_{i+1})^2 = 3\,P^{(i)}_0 + \mathbf{1}$, where $P_0^{(i)}$ is the projector into total spin zero for the neighboring sites.
Dropping the additive constant and defining
$e_i \equiv 3\, P_0^{(i)}$, the pure biquadratic
Hamiltonian takes the form $H = \sum_i e_i$.
The generators $e_i$ so defined satisfy three relations:
$e_i^2 = \delta\, e_i$ with $\delta = 2s+1$
(equal to~3 for spin-1, since
$e_i^2 = 9\, P_0 = 3\, e_i$),
$e_i\, e_{i\pm 1}\, e_i = e_i$ (the braid-like relation
that couples neighboring bonds), and
$[e_i, e_j] = 0$ for $|i - j| \geq 2$ (locality).
These are the defining relations of the
Temperley--Lieb (TL) algebra~\cite{Temperley:1971iq},
with loop fugacity $\delta = 3$.

The TL algebra endows the biquadratic chain with both
a symmetry structure and integrability.
On the symmetry side, because the Hamiltonian is built
entirely from TL generators, the TL algebra---rather
than $SU(2)$---is the relevant algebraic structure
organizing the spectrum.
The TL algebra is also intimately connected to the
quantum group $U_q(sl_2)$ via
$q + q^{-1} = \delta$~\cite{Chari:1994pz,Pasquier:1989kd},
linking it to the root-of-unity discussion in
Sec.~\ref{sec:root_of_unity}, although here $q$ is
real and the representation theory remains semisimple.
On the integrability side, from the TL generators one
can construct an $R$-matrix satisfying the Yang--Baxter
equation, making the pure biquadratic chain exactly
solvable by Bethe ansatz~\cite{Barber:1989fz,Kluemper:1989}.

The contrast with the Takhtajan--Babujian point is
instructive: both $\theta = \pi/2$ and
$\theta = -\pi/4$ are integrable and Bethe-ansatz
solvable~\cite{Takhtajan:1982,Babujian:1982,Barber:1989fz},
yet only the biquadratic point exhibits a dip
in~$\avg{\ep}$.
The crucial difference lies  in
the spectral degeneracy pattern imposed by the TL
algebra.
The Hilbert space $(\CC^3)^{\otimes L}$ decomposes
into irreducible TL modules whose dimensions are
determined by Chebyshev polynomials of the second kind
evaluated at
$\delta/2$~\cite{Martin:1991xx}.
Concretely, for $L = 2$ the TL decomposition yields
the degeneracy pattern
$\{1, 1, 1, 1, 1, 1, 1, 1, 4\}$ noted above---eight
singlet blocks and one four-dimensional block---which
is far more fragmented than the $SU(2)$ pattern
$\{1, 3, 5\}$.
For larger $L$ the disparity grows rapidly: the TL
modules produce degenerate blocks that are much larger
than those of the standard $SU(2)$ or $SU(3)$
multiplet decompositions.
It is this extreme TL degeneracy, rather than
integrability or any continuous symmetry, that drives
the strong suppression of~$\avg{\ep}$ at the
biquadratic point.

\section{Thermodynamic Limit: Quasi-particle Scatterings}
\label{sec:twomagnon}

The results of the preceding sections demonstrate that
the many-body $\avg{\ep}$ is a sensitive probe of
symmetry-induced spectral degeneracies in finite
systems.
Integrability can amplify these degeneracies---as at
the free-fermion and root-of-unity points---but need
not do so: the Takhtajan--Babujian point is
Bethe-ansatz integrable yet produces no discernible
dip.
The entangling power therefore detects algebraic
structure that manifests as enhanced degeneracies,
regardless of whether that structure originates from
a continuous symmetry, a quantum group, or the
Temperley--Lieb algebra.

It is then natural that the signals in the
entangling power decay with system
size---the Hilbert-space dimension grows
exponentially with $L$, while the number of
constraints grows only polynomially,
so their relative weight is progressively diluted.
Far from being a shortcoming, this vanishing is a
reflection of the fact that $\avg{\ep}$ probes the
global structure of the {\em entire} spectrum and
its eigenvectors, a structure that becomes
increasingly dominated by generic, symmetry-agnostic
features as the system grows.

To construct a diagnostic that survives the
thermodynamic limit, we must therefore shift focus
from the full many-body spectrum to the low-lying
excitations whose structure is most tightly
constrained by symmetry.
The idea that quasi-particle degrees of freedom
provide the natural language for entanglement in
the thermodynamic limit has a distinguished pedigree:
Calabrese and Cardy showed that the entanglement
entropy of a one-dimensional quantum system after a
global quench can be understood in terms of entangled
quasi-particle pairs propagating ballistically from
the initial
state~\cite{Calabrese:2004eu,Calabrese:2009qy,Alba:2017pnas}.
In their picture, quasi-particles are the {\em carriers}
of entanglement across the system.
Here we apply a complementary philosophy: rather than
asking how quasi-particles propagate pre-existing
entanglement, we ask how much entanglement is
{\em generated} by the dynamics of  quasi-particles, which should inherit the dynamics of the underlying spin-chains.

This viewpoint is closely related to the philosophy
of effective field theory: the full
many-body Hilbert space, with its exponentially
growing dimension, plays the role of the UV theory,
while the quasi-particle description provides the
low-energy effective theory whose dynamics are
governed by the symmetries of the underlying system.

Concretely, it is worth recalling that the correlation between
entanglement suppression and symmetry enhancement
was first discovered in $2 \to 2$ scattering
amplitudes in particle and nuclear
physics~\cite{Beane:2018oxh,Low:2021ufv}, where the
$S$-matrix is the fundamental observable.
Two-magnon scattering is the spin-chain counterpart
of this setting: magnons are the elementary
excitations, their pairwise $S$-matrix is fixed by
the Yang--Baxter equation, and its entangling power
provides an intensive, per-scattering-event
diagnostic of symmetry that persists as
$L \to \infty$.
In this way, the quasi-particle picture brings the
many-body problem back into contact with the
scattering framework in which the
entanglement--symmetry connection was originally
identified.

\subsection{Two-magnon $S$-matrix}
\label{sec:6vertex}

We now switch from open to periodic boundary conditions,
replacing the sum in Eq.~(\ref{eq:H_chain}) by
$\sum_{j=1}^{L}$ with $\vec{S}_{L+1} \equiv \vec{S}_1$.
The periodic XXZ Hamiltonian is exactly solvable by the
coordinate Bethe
ansatz~\cite{Bethe:1931hc,Orbach:1958,Yang:1966ek,desCloizeaux:1962}.
(The finite-size calculations of Secs.~\ref{sec:finitesize}
and~\ref{sec:integrability} employed open boundary conditions,
which are natural for the quantum group analysis; the Bethe ansatz,
by contrast, requires translational invariance.)
The basic idea is simple: one works in the sector
of fixed total $S^z$ and writes an ansatz for the
wavefunction as a superposition of plane waves,
one for each ordering of the flipped spins.
Consider the ferromagnetic vacuum
$|\Omega\rangle = |\!\uparrow\uparrow
\cdots\uparrow\rangle$, the state with all spins
aligned.
Spin-flip excitations above this vacuum are called
magnons.
A single magnon is created by flipping one spin:
acting with $S^-_n$ on site $n$ produces a localized
excitation $|n\rangle = S^-_n|\Omega\rangle$.
In the Bethe ansatz, the one-magnon eigenstate is
a Bloch wave,
$|p\rangle = \sum_{n=1}^{L} e^{ipn}\,|n\rangle$,
carrying quasi-momentum $p$ and spin
$\Delta S^z = -1$, with dispersion relation
$\epsilon(p) = \Delta - \cos p$. (A self-contained derivation is given in
Appendix~\ref{app:bethe}.)
A single magnon thus propagates freely;
when two magnons meet on adjacent sites, however,
the wavefunction acquires a phase shift.
This phase shift depends on the anisotropy $\Delta$
and on the {\em rapidity} difference of the two
magnons.
The rapidity is a reparametrization of the
quasi-momentum that linearizes the scattering
phase---it plays the same role here as the rapidity
variable in relativistic kinematics, trading a
complicated dependence on momenta for a simple
dependence on their difference.

The two-magnon scattering is encoded in the
$R$-matrix, which is the scattering amplitude in
spin and rapidity space: it maps an incoming
two-magnon spin state to an outgoing one as a
function of the rapidity difference.
Since the XXZ Hamiltonian conserves total $S^z$,
the scattering cannot change the total spin
projection of the two magnons.
Written in the basis
$\{|\!\uparrow\uparrow\rangle,\,
|\!\uparrow\downarrow\rangle,\,
|\!\downarrow\uparrow\rangle,\,
|\!\downarrow\downarrow\rangle\}$,
this means that any process connecting states with
different total $S^z$---such as
$|\!\uparrow\uparrow\rangle \to
|\!\uparrow\downarrow\rangle$---has vanishing
amplitude.
Of the sixteen entries, only six survive, giving
rise to the so-called six-vertex
form~\cite{Lieb:1967zz,Lieb:1961fr,Baxter:1972hz}:
\be
\label{eq:R_6v}
R(u,\gamma) = \begin{pmatrix}
a & 0 & 0 & 0 \\
0 & b & c & 0 \\
0 & c & b & 0 \\
0 & 0 & 0 & a
\end{pmatrix}\,,
\ee
with vertex weights $a = \sin(u + \gamma)$,
$b = \sin u$, $c = \sin\gamma$, and
$\Delta = \cos\gamma$~\cite{Baxter:1972hz}.
Here $u$ is the spectral parameter encoding the
rapidity difference---it is related to the
quasi-momentum~$p$ of each magnon by
$e^{ip} = \sin(u + \gamma/2)/\sin(u - \gamma/2)$
(see Appendix~\ref{app:rapidity} for details)---and
$\gamma$ is the crossing parameter fixed by the
anisotropy.
The three types of nonzero entries correspond to
the three allowed scattering processes:
$a$ is the amplitude for two magnons with the same
spin to pass through each other unchanged
($S^z = \pm 1$ sectors),
$b$ is the amplitude for opposite-spin magnons to
transmit without exchanging spin,
and $c$ is the spin-exchange amplitude---the process
$|\!\uparrow\downarrow\rangle
\leftrightarrow
|\!\downarrow\uparrow\rangle$.
It is the spin-exchange amplitude $c$ that generates
entanglement; when $c = 0$ the $R$-matrix is
diagonal and no entanglement is produced.

A key consequence of integrability is that the
two-magnon scattering is purely elastic---the magnons
emerge with their momenta unchanged, picking up only
a phase---and that the multi-magnon $S$-matrix
factorizes into a product of pairwise two-body
scatterings~\cite{Zamolodchikov:1978xm}.
The self-consistency of this factorization is
guaranteed by the Yang--Baxter
equation~\cite{Yang:1967bm,Baxter:1972hz,Sklyanin:1979},
\begin{multline}
\label{eq:ybe_main}
R_{12}(u_{12})\, R_{13}(u_{13})\, R_{23}(u_{23})  \\
= R_{23}(u_{23})\, R_{13}(u_{13})\, R_{12}(u_{12})\,,
\end{multline}
where $u_i$ is the rapidity of magnon $i$,
$u_{ij} \equiv u_i - u_j$, and $R_{jk}$ denotes the
$R$-matrix of Eq.~(\ref{eq:R_6v}) acting on magnons
$j$ and~$k$.
The crossing parameter $\gamma$ is common to all
$R$-matrices in the chain, so it is suppressed; only
the rapidity difference $u_{ij}$ varies from factor
to factor.
Eq.~(\ref{eq:ybe_main}) is a matrix identity in
$\CC^2 \otimes \CC^2 \otimes \CC^2$:
it states that the outcome of a three-magnon
collision is independent of the order in which the
pairwise scatterings are performed---the hallmark
of an integrable system.

The $R$-matrix as written above encodes the
relative scattering amplitudes but is not yet a
proper $S$-matrix, because it is not unitary---the
outgoing probabilities do not sum to one.
This is an artifact of the parametrization: the
spectral parameter $u$ is natural in the algebraic
setting of the vertex model, but physical
scattering corresponds to real rapidity difference
$\theta$, related by $u = i\theta$.
Making this substitution and dividing by an overall
normalization
$\rho = \sqrt{\sin^2\!\gamma + \sinh^2\!\theta}$
to enforce unitarity, we obtain the physical
$S$-matrix,
$S(\theta,\gamma) = R(i\theta,\gamma)/\rho$,
which is unitary by construction and describes
the physical scattering of two magnons
(see Appendix~\ref{app:bethe} for details).

\subsection{Entangling power of the $S$-matrix}
\label{sec:ep_smatrix}

Since $S(\theta, \gamma)$ conserves total $S^z$,
its action on $\CC^2 \otimes \CC^2$ decomposes
into three sectors.
In the $S^z = \pm 1$ sectors
($|\!\!\uparrow\uparrow\rangle$ and
$|\!\!\downarrow\downarrow\rangle$), no spin exchange
is possible and the $S$-matrix acts as a phase:
$S = e^{i\varphi_+}$.
In the $S^z = 0$ sector, spanned by
$|\!\uparrow\downarrow\rangle$ and
$|\!\downarrow\uparrow\rangle$, the $2\times 2$ block
has two eigenvalues corresponding to the triplet
($e^{i\varphi_t}$, symmetric) and singlet
($e^{i\varphi_s}$, antisymmetric) channels.
Explicitly, the three phases are
\begin{align}
e^{i\varphi_+} &= a/\rho\,,  \notag \\
e^{i\varphi_t} &= (b + c)/\rho\,, \notag \\
e^{i\varphi_s} &= (b - c)/\rho\,,
\label{eq:smat_phases}
\end{align}
where, after the analytic continuation
$u = i\theta$ to physical rapidity,
\begin{align}
a &= \sin\gamma\,\cosh\theta
+ i\cos\gamma\,\sinh\theta\,, \notag \\
b &= i\sinh\theta\,, \qquad
c \;=\; \sin\gamma\,,
\label{eq:abc_physical}
\end{align}
with $\rho = \sqrt{\sin^2\!\gamma + \sinh^2\!\theta}$
and $\Delta = \cos\gamma$.

At fixed rapidity
$\theta$, the $S$-matrix is a finite-dimensional
unitary on the spin Hilbert space
$\CC^2 \otimes \CC^2$---the rapidity enters as a
continuous parameter. In the spin-subspace
the entangling power of $S(\theta, \gamma)$ is
therefore that of a two-qubit gate, and it depends
only on the eigenvalue phases.
For the $S^z$-conserving structure above, the
overall phase $\varphi_+$ drops out, and
$\ep$ is a function of the two phase differences
$\delta_{t+} \equiv \varphi_t - \varphi_+$ and
$\delta_{s+} \equiv \varphi_s - \varphi_+$ alone.

To derive the explicit formula, we note that for
two qubits ($d_A = d_B = 2$) the linear entropy
simplifies to $1 - \tr(\rho_A^2) = 2\det\rho_A$,
so the entangling power in Eq.~(\ref{eq:ep_haar})
reduces to twice the Haar average of the determinant of the
reduced density matrix over random product input
states.
We parametrize the two input states on their
respective Bloch spheres, exploiting the $U(1)$
symmetry to fix one azimuthal angle.

Analytically, the key simplification is that the
$S$-matrix conserves $S^z$ and therefore has
only a transmission amplitude $B$ and a
spin-exchange amplitude $C$ in the $S^z = 0$
sector, with $|B|^2 + |C|^2 = 1$.
When computing $\det\rho_A$ and integrating
over the remaining azimuthal angle, all
cross-terms between $B$ and $C$ contain oscillating
phases and vanish, leaving two decoupled
contributions that can be evaluated in closed form
(see Appendix~\ref{app:bethe} for the detailed
derivation).
The result is
\begin{multline}
\label{eq:ep_main}
\ep(\delta_{t+}, \delta_{s+}) = \frac{1}{18}\bigl[
3 - 2\cos(\delta_{t+} + \delta_{s+})\cos(\delta_{t+} - \delta_{s+})  \\
- \cos^2(\delta_{t+} - \delta_{s+})\bigr]\,.
\end{multline}
Eq.~(\ref{eq:ep_main}) is the central formula
of this section: it expresses the entangling
power of any
$S^z$-conserving two-qubit unitary entirely in
terms of two phase differences, with no reference
to the specific dynamical model that produced them.

\subsection{Entanglement  and Symmetry}

It is instructive to decompose the $S$-matrix
in terms of quantum logic gates.
By direct inspection one can write
\be
\label{eq:smat_decomp}
S(\theta,\gamma) = \tfrac{1}{2}(a{+}b{-}c)\, I
\;+\; c\, {\rm SWAP}
\;+\; \tfrac{1}{2}(a{-}b{-}c)\,
\sigma_z \otimes \sigma_z\,,
\ee
which is the $U(1)$ generalization of the two-gate
decomposition in
Ref.~\cite{Low:2021ufv}.
Under $SU(2)$ invariance, $\sigma_z\!\otimes\!\sigma_z$ is
forbidden and the third term must vanish, giving
$a = b + c$ and $S = b\,I + c\,{\rm SWAP}$,
recovering the nucleon-nucleon result.
The $\sigma_z\!\otimes\!\sigma_z$ coefficient
$\tfrac{1}{2}(a - b - c)$
measures the departure from $SU(2)$: it reflects
the anisotropy-induced splitting of the triplet
eigenvalues ($\varphi_+ \neq \varphi_t$),
permitted by the reduced $U(1)$ symmetry of the
XXZ chain.

The key insight from Ref.~\cite{Low:2021ufv} is
that the Identity and the SWAP are the \emph{only}
two equivalence classes of two-qubit gates
with vanishing entanglement
power~\cite{Zhang:2003qph,Balakrishnan:2010epli}.
(Here $\sigma_z\!\otimes\!\sigma_z$ is a product
of local unitaries and hence belongs to the
Identity class.)
In the $(\theta, \Delta)$ plane, $\ep = 0$
therefore requires $S$ to be locally equivalent
to one of these two gates.
From Eq.~(\ref{eq:smat_decomp}), this happens
in exactly two regimes:

(i)~\emph{$SU(2)$ points}
($\Delta = \pm 1$, all $\theta$).
The spin-exchange amplitude vanishes,
$c = \sin\gamma = 0$, and one can verify from
Eq.~(\ref{eq:abc_physical}) that $a = b$ at
$\gamma = 0$ and $a = -b$ at $\gamma = \pi$.
In both cases only a single term in
Eq.~(\ref{eq:smat_decomp}) survives---proportional
to $I$ or to $\sigma_z\!\otimes\!\sigma_z$---and
the $S$-matrix belongs to the Identity class
for \emph{all}~$\theta$.
The rapidity independence has a simple
physical origin: at the $SU(2)$ point
the $S$-matrix acts trivially on spin,
so the outgoing two-magnon state is an unentangled product state
of spin and momentum regardless of the
scattering energy.
This is the spin-chain counterpart
of the entanglement suppression found at the
Wigner $SU(4)$ point in nucleon-nucleon
scattering,
where the $S$-matrix in the spin-subspace is again an Identity, independent of the centre-of-mass momentum~\cite{Low:2021ufv}.

(ii)~\emph{Zero rapidity}
($\theta = 0$, $\Delta \neq \pm 1$).
At zero relative momentum
$a = c = \sin\gamma$ and $b = 0$, so
$S \propto {\rm SWAP}$: the two magnons exchange
their spin quantum numbers completely.
Since SWAP is the other non-entangling gate,
$\ep = 0$ here as well, but for a different
reason---SWAP rather than Identity.

\begin{figure}[t]
\includegraphics[width=\columnwidth]{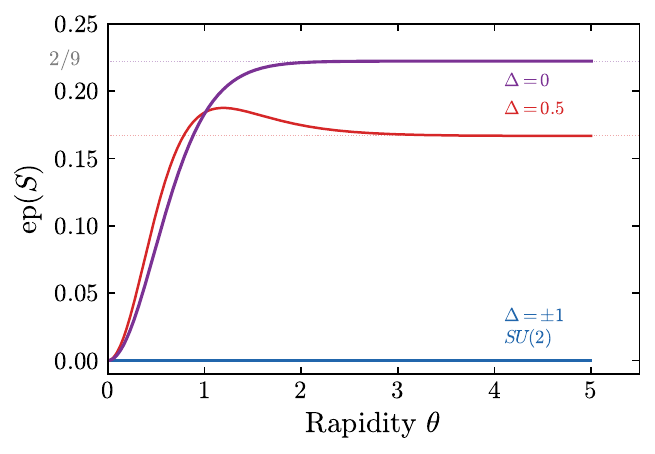}
\caption{\label{fig:ep_smatrix}
{\em Entangling power of the two-magnon
$S$-matrix as a function of rapidity $\theta$
for several values of the anisotropy $\Delta$.}
At the $SU(2)$ points $\Delta = \pm 1$,
$S$ belongs to the Identity class and
$\ep = 0$ for all $\theta$ (selection rule).
For $\Delta \neq \pm 1$, $\ep$ rises from
zero ($S \propto {\rm SWAP}$ at $\theta = 0$)
and approaches the asymptotic value
$\ep_\infty = \tfrac{2}{9}(1 - \Delta^2)$
(dotted lines).
The free-fermion point $\Delta = 0$ achieves the
maximum at every $\theta$.
}
\end{figure}

Away from these two regimes, all three terms in
Eq.~(\ref{eq:smat_decomp}) contribute, the
$S$-matrix belongs to neither non-entangling
class, and $\ep > 0$.
As the rapidity increases from zero, $S$
interpolates away from the SWAP gate and
entanglement is generated.
At fixed $\theta$, smaller $|\Delta|$ corresponds
to a larger spin-exchange amplitude
$c = \sqrt{1 - \Delta^2}$, and hence larger $\ep$.
The maximum is achieved at the free-fermion point
$\Delta = 0$ ($c = 1$), where the $S$-matrix has
maximal spin exchange and $\ep$ approaches the
two-qubit ceiling $2/9$ at large rapidity
(Fig.~\ref{fig:ep_smatrix}).
It is worth pointing out the contrast with the
full-spectrum results of the previous sections,
where $\Delta = 0$ corresponded to a sharp
\emph{dip} in entangling power:
there, the free-fermion structure simplified the
many-body eigenstates and suppressed entanglement
across the full Hilbert space, whereas the
quasi-particle $S$-matrix at $\Delta = 0$ has
the \emph{largest} spin-exchange amplitude and is
therefore the most entangling.
These are complementary perspectives---the full
spectrum probes the global structure of the
Hilbert space, while the $S$-matrix isolates
the local two-body interaction.
It is also worth noting that entanglement
suppression is not the only correlation between
entanglement and symmetry: in the two-Higgs-doublet
model, Ref.~\cite{Carena:2025wyh} showed that
\emph{maximizing} the entanglement  in
Higgs boson scattering also results in enhanced symmetry,
providing a complementary example where enhanced
symmetry seems to correspond to the extremum in the entanglement, instead of just the minimum.

In the large-rapidity limit
$\sinh\theta \gg \sin\gamma$, both phase differences
approach the same value,
$\delta_{t+}, \delta_{s+} \to \gamma$, and the $S$-matrix reduces
to the Ising-type $ZZ$ rotation
$e^{-i\gamma\,\sigma_z\otimes\sigma_z/2}$.
Eq.~(\ref{eq:ep_main}) then gives the asymptotic
entangling power
\be
\label{eq:ep_inf}
\ep_\infty(\gamma) = \frac{2}{9}\sin^2\!\gamma
= \frac{2}{9}(1 - \Delta^2)\,,
\ee
which provides a simple closed-form expression
for the dotted asymptotes in
Fig.~\ref{fig:ep_smatrix}.

\section{Conclusions}
\label{sec:discussion}

In this work we have undertaken a systematic study of
the entangling power of spin-chain Hamiltonians,
encompassing two-site models, finite-size chains,
and the two-magnon scattering matrix.
The central finding is that the correlation between
entanglement suppression and symmetry enhancement,
previously observed in particle and nuclear physics,
extends to spin-chain systems across all settings we
have studied.

The exact solution of the two-site models yields an
insight that is invisible in $2 \to 2$ scattering
amplitudes in particle and nuclear
physics~\cite{Beane:2018oxh,Low:2021ufv}: the
time-averaged entangling power depends on
\emph{both} the eigenvalue spectrum and the
eigenvector structure of the Hamiltonian.
For spin-1/2, the eigenvectors are universal
Bell states independent of the couplings, so
$\avg{\ep}$ is determined entirely by the eigenvalue
degeneracy pattern---the count $N_0$ of vanishing
frequencies suffices.
For spin-1, the eigenvectors become
coupling-dependent, and the isospectral analysis
reveals that the eigenvector contribution to the
suppression at the XXX point is comparable in
magnitude to the eigenvalue contribution.
This dual mechanism---eigenvalue degeneracy and
eigenvector constraint---is a structural feature of
many-body Hamiltonians that has no counterpart in
the $S$-matrix, where only eigenvalue phases enter
the entangling power.
We established the monotonic
hierarchy $\avg{\ep}_{\rm XYZ} > \avg{\ep}_{\rm XXZ}
\geq \avg{\ep}_{\rm XX} > \avg{\ep}_{\rm Ising}
> \avg{\ep}_{\rm XXX}$
for both spin-1/2 and spin-1.

For finite-size systems, the symmetry dip persists
but its relative depth decays with increasing~$L$.
This is not a shortcoming of the entangling power
but a reflection of the fact that it probes
the \emph{entire} eigenspace---all $d^4$ quartets of
eigenstates contribute to the time-averaged purity
sums.
As $L$ grows, the Hilbert space dimension $d$ grows
exponentially while the number of symmetry-imposed
constraints grows only polynomially; the symmetry
signal is progressively diluted.
Meanwhile, the free-fermion dip at $\Delta = 0$
decays much more slowly, driven by the additivity of the
Jordan--Wigner spectrum---a mechanism invisible to
any finite-body scattering amplitude.
Integrability breaking via a $J_2$ coupling confirms
that symmetry, not integrability, is the driver of
the $SU(2)$ suppression.
The bilinear-biquadratic spin-$1$ chain provides a
complementary lesson: the $SU(3)$-symmetric ULS point
produces less pronounced
suppression than the biquadratic point, while the pure biquadratic
point---whose extreme Temperley--Lieb degeneracy far
exceeds the $SU(2)$ or $SU(3)$ multiplet structure---hosts
the deepest dip of any model we have studied.
This confirms that it is the detailed spectral degeneracy
pattern, not merely the dimension of the symmetry group,
that controls the suppression of~$\avg{\ep}$.
We note that for the free-fermion and root-of-unity dips,
the mechanism---polynomial growth of distinct energy
differences and indecomposable quantum-group
representations, respectively---identifies {\em why}
large $\omega$-groups form, but the quantitative dip
depths are established numerically rather than
analytically; closing this gap is an interesting
open problem.

That the many-body entangling power fades with
system size points naturally to the correct
diagnostic in the thermodynamic limit: the dynamics
of quasi-particles.
Two-magnon scattering is the spin-chain counterpart
of $2\to 2$ scattering in particle physics, and its
entangling power is an intensive, per-scattering-event
quantity that persists as $L \to \infty$.
We showed that the $S$-matrix decomposes into three quantum logic
gates---Identity, SWAP, and
$\sigma_z\!\otimes\!\sigma_z$---generalizing the
$SU(2)$ two-gate decomposition of
Ref.~\cite{Low:2021ufv} to the $U(1)$-symmetric
XXZ chain.
The key insight, also from Ref.~\cite{Low:2021ufv},
is that Identity and SWAP are the \emph{only}
non-entangling two-qubit gates.
The entangling power of the $S$-matrix therefore
vanishes if and only if it belongs to one of these
two classes, which occurs at the $SU(2)$ points
$\Delta = \pm 1$ for all scattering energies.
The free-fermion point $\Delta = 0$ achieves the
maximum entangling power at every rapidity, in
striking contrast to  the full-spectrum result
where it corresponds to a sharp dip---illustrating
that the $S$-matrix and the many-body spectrum
probe complementary aspects of entanglement.

Taken together, these results trace a single thread
across three regimes connected by very different
physics: suppressed entangling power at points of
enhanced algebraic structure, from microscopic quantum
gates through finite-size many-body spectra to
asymptotic two-body scattering.
That the correlation persists across regimes---and
is driven in each case by the spectral degeneracies
and eigenvector constraints imposed by the
symmetry---suggests it is a robust organizing
principle rather than an artifact of any single
setting.

It is worth pointing out that the finite-size entanglement
power is directly relevant to near-term quantum platforms.
Programmable quantum simulators based on cold atoms,
trapped ions, or superconducting
qubits~\cite{Bloch:2012bh,Gross:2017hnh,Monroe:2021rmp}
now routinely
realize spin chains with tunable anisotropy in the range
of system sizes we have studied ($L = 4$--$10$).
The dip structure of the time-averaged entangling power
as a function of the anisotropy parameter could serve as
a symmetry witness: one prepares random product states,
evolves under the implemented Hamiltonian, and measures
the average entanglement produced.
A key advantage of this protocol is that it probes
the symmetry content of the dynamics itself, without
requiring the preparation of a specific eigenstate or
{\em a priori} knowledge of the conserved charges.
This distinguishes it from state-based diagnostics
such as ground-state entanglement entropy or
symmetry-resolved entanglement, which presuppose
both.
Deviations from the predicted dip pattern would signal
that hardware noise or calibration errors have broken the
intended symmetry of the simulation.
More broadly, the entangling power generalizes the
standard two-qubit gate characterization to multi-qubit
unitaries, which is of interest for benchmarking
entangling blocks in quantum circuit design.

Several directions for future work suggest themselves.
A quantitative connection between the $S$-matrix
entangling power and the many-body dip depth
remains to be established.
It would also be interesting to extend the gate
decomposition to higher-rank models
($d = 3$ and beyond) where the $R$-matrix is known.
Finally, spectral correlations beyond the entanglement
power---such as the cross spectral form factor introduced
in Ref.~\cite{Bai:2026xsff} for bootstrapping hidden
symmetries from level statistics---offer a complementary route
to diagnosing symmetry and integrability in many-body
systems, and understanding the interplay between these
spectral approaches remains an open challenge.

\section*{Acknowledgments}
We thank Daniel Arovas, Jens Koch and Shu-Heng Shao 
for discussions. This work is supported in part by the U.S. Department of Energy under contracts No. DE-SC0023522, No. DE-SC0010143, and No.
89243024CSC000002 (QuantISED Program). This research was supported in part through the computational resources and staff contributions provided for the Quest high performance computing facility at Northwestern University which is jointly supported by the Office of the Provost, the Office for Research, and Northwestern University Information Technology.

The AI assistant Claude (Anthropic) was used during this work for code implementation and manuscript preparation. Code used in this work was produced with AI assistance under close human supervision and was validated against known analytical results and internal consistency tests. The authors originated the scientific ideas and direction of the work, developed the physical insights and logical flow of the paper, and extensively revised, restructured, and finalized the manuscript text. The authors take full responsibility for the accuracy, originality, and integrity of the work, and no AI tool is listed as an author.

\appendix

\section{Derivation of the time-averaging algorithm in Section \ref{subsect:time}}
\label{app:algorithm}

In this appendix we derive the algorithm presented in
Sec.~\ref{sec:ep} for computing the time-averaged
entangling power $\avg{\ep}$.
Throughout we specialize to the equal bipartition
$d_A = d_B \equiv d$, which is the case relevant for
the spin-chain calculations in this work.
The starting point is the operator
formula~(\ref{eq:Ialpha}), which we write for
$\alpha = 0$ as
\bea
\label{eq:I0_decomp}
I_0(U) &=& \tr(T_{13})
\nonumber\\
&& +\; \tr\!\big[(U^\dagger)^{\otimes 2}\,
   T_{13}\, U^{\otimes 2}\, T_{13}\big]\,.
\eea
For a time-independent Hamiltonian,
$U(t) = e^{iHt} = \sum_n e^{iE_n t}|n\>\<n|$, and in
the doubled Hilbert space
$({\cal H}_A\otimes{\cal H}_B)^{\otimes 2}$ we have
$U^{\otimes 2}(t) = \sum_{k,l} e^{i(E_k+E_l)t}\,
|kl\>\<kl|$,
where $|kl\> \equiv |k\>_1\otimes|l\>_2$ labels
the energy eigenstates in the first and second copies.
Inserting this into Eq.~(\ref{eq:I0_decomp}) and using
$T_{13}^\dagger = T_{13}$, we obtain
\bea
\label{eq:I0_expanded}
I_0^{(2)}(t) &\equiv&
 \tr\!\big[(U^\dagger)^{\otimes 2}\,T_{13}\,
   U^{\otimes 2}\,T_{13}\big]\nonumber\\
 &=& \sum_{k,l,m,n} e^{i(E_k+E_l-E_m-E_n)\,t}
\nonumber\\
 && \qquad\times\;
   \big|\<kl|T_{13}|mn\>\big|^2\,.
\eea
The modulus squared arises because $T_{13}$ is
Hermitian and appears on both sides of $U^{\otimes 2}$:
$\<mn|T_{13}|kl\> = (\<kl|T_{13}|mn\>)^*$.

\subsection{Matrix element of the swap operator}
\label{app:matrix_element}

We now evaluate $\<kl|T_{13}|mn\>$ in the product basis.
Let $\{|a\>\}$ and $\{|b\>\}$ be orthonormal bases for
${\cal H}_A$ and ${\cal H}_B$, respectively.
Each eigenvector decomposes as
$|n\> = \sum_{a,b}(C_n)_{ab}\,|a\>\otimes|b\>$,
where $C_n$ is a $d_A\times d_B$ matrix.
In the doubled space the basis is
$|a_1,b_1,a_2,b_2\>$, with positions 1 and 2
labelling the $A$-factor and the $B$-factor of each
copy.
The swap operator $T_{13}$ exchanges the two $A$-factors:
\be
T_{13}\,|a_1,b_1,a_2,b_2\> = |a_2,b_1,a_1,b_2\>\,.
\ee
Therefore
\bea
&&\<kl|T_{13}|mn\>
\nonumber\\
 &=& \sum_{a,b,c,d}
   (C_k)^*_{cb}\,(C_l)^*_{ad}\,
   (C_m)_{ab}\,(C_n)_{cd}
   \nonumber\\
 &=& \sum_{a,c}\Big(\sum_b (C_m)_{ab}\,
   (C_k^\dagger)_{bc}\Big)
\nonumber\\
 && \qquad\times\;
   \Big(\sum_d (C_n)_{cd}\,
   (C_l^\dagger)_{da}\Big)
   \nonumber\\
 &=& \tr_A(C_m C_k^\dagger\cdot C_n C_l^\dagger)\,.
\label{eq:T13_me}
\eea
In the first line we expanded $|mn\>$ and $\<kl|$,
applied $T_{13}$ (which sends $a_1\to a_2$ and
$a_2\to a_1$), and contracted the indices.
The second line collects the sums over $b$ and $d$
into matrix products, yielding the trace over
${\cal H}_A$ of two $d_A\times d_A$ matrices.

\subsection{Time average and eigenvalue grouping}
\label{app:time_avg}

The infinite-time average of Eq.~(\ref{eq:I0_expanded})
retains only those quartets $(k,l,m,n)$ satisfying
$E_k + E_l = E_m + E_n$, or equivalently
$E_k - E_m = E_n - E_l \equiv \omega$:
\bea
\label{eq:I0_avg}
\avg{I_0^{(2)}} &=& \sum_\omega\;
 \sum_{\substack{(k,m):\,E_k-E_m=\omega\\
                 (n,l):\,E_n-E_l=\omega}}
\nonumber\\
&& \qquad\times\;
 \big|\tr_A(C_m C_k^\dagger \cdot
  C_n C_l^\dagger)\big|^2\,.
\eea
For each value of $\omega$, define a group $g_\omega$
consisting of all pairs of eigenstates $(p,q)$ with
$E_p - E_q = \omega$, and label them
$i = 1,2,\ldots,N_\omega$.
For the $i$th pair $(p_i,q_i)$, define the
$d_A \times d_A$ matrix
\be
M_i \equiv C_{p_i}\,C_{q_i}^\dagger\,.
\ee
Using Eq.~(\ref{eq:T13_me}), the matrix element
becomes
$\tr_A(C_m C_k^\dagger \cdot C_n C_l^\dagger)
 = \tr_A(M_i^\dagger M_j)$
where $i$ labels the pair $(k,m)$ and $j$ labels the
pair $(n,l)$, both belonging to the same group
$g_\omega$.
Equation~(\ref{eq:I0_avg}) then takes the compact form
\be
\label{eq:I0_grouped}
\avg{I_0^{(2)}} = \sum_\omega
 \sum_{i,j=1}^{N_\omega}
 \big|\tr_A(M_i^\dagger M_j)\big|^2\,.
\ee

\subsection{Relation between $I_1$ and $I_0$}
\label{app:I1}

The formula for $\alpha=1$ reads
\bea
I_1(U) &=& \tr(T_{24})
\nonumber\\
&& +\; \tr\!\big[(U^\dagger)^{\otimes 2}\,
   T_{24}\, U^{\otimes 2}\, T_{13}\big]\,.
\eea
Unlike $I_0$, the two swap operators flanking
$U^{\otimes 2}$ are different ($T_{24}$ and $T_{13}$),
so the $|\cdot|^2$ factorization of
Eq.~(\ref{eq:I0_expanded}) does not apply directly.
Instead, we relate $I_1(U)$ to $I_0$ evaluated on a
modified unitary.

Let $S$ denote the SWAP gate on
${\cal H}_A\otimes{\cal H}_B$, i.e.\
$S|a,b\> = |b,a\>$.
We claim that $S^{\otimes 2}\,T_{13}\,S^{\otimes 2}
= T_{24}$.
This is verified by acting on a basis vector:
\bea
&&|a_1,b_1,a_2,b_2\>
 \xrightarrow{S^{\otimes 2}} |b_1,a_1,b_2,a_2\>
 \nonumber\\
&&\quad\xrightarrow{T_{13}} |b_2,a_1,b_1,a_2\>
 \xrightarrow{S^{\otimes 2}} |a_1,b_2,a_2,b_1\>\,,
 \nonumber
\eea
which is precisely the action of $T_{24}$.
Since $S^\dagger = S$ and $S^2 = \Id$, we compute
\bea
I_0(SU) &=& \tr(T_{13})
 + \tr\!\big[(U^\dagger)^{\otimes 2}\,
   S^{\otimes 2} T_{13} S^{\otimes 2}\,
   U^{\otimes 2}\, T_{13}\big]
\nonumber\\
 &=& \tr(T_{13})
 + \tr\!\big[(U^\dagger)^{\otimes 2}\,
   T_{24}\, U^{\otimes 2}\, T_{13}\big]\,.
\label{eq:I0SU}
\eea
Comparing with $I_1(U) = \tr(T_{24})
+ \tr[(U^\dagger)^{\otimes 2}\,T_{24}\,
U^{\otimes 2}\,T_{13}]$,
we see that $I_0(SU)$ and $I_1(U)$ share the same
dynamical term and differ only in their constant
parts: $\tr(T_{13})$ versus $\tr(T_{24})$.
For $d_A = d_B = d$ (as is always the case in the
spin chains considered in this work),
$\tr(T_{13}) = \tr(T_{24}) = d^3$, and therefore
\be
\label{eq:I1_I0SU}
I_1(U) = I_0(SU)\qquad (d_A = d_B)\,.
\ee
To derive the algorithmic form of $\avg{I_1^{(2)}}$,
we evaluate $\<kl|T_{24}|mn\>$ directly.
The swap operator $T_{24}$ exchanges the two
$B$-factors:
$T_{24}\,|a_1,b_1,a_2,b_2\> = |a_1,b_2,a_2,b_1\>$.
Repeating the index contraction of
Sec.~\ref{app:matrix_element} with this exchange, we
find
\bea
\<kl|T_{24}|mn\>
 &=& \sum_{\substack{a_1,b_1\\a_2,b_2}}
   (C_k)^*_{a_1 b_2}\,(C_l)^*_{a_2 b_1}\,
   (C_m)_{a_1 b_1}\,(C_n)_{a_2 b_2}
   \nonumber\\
 &=& \tr_B(C_k^\dagger C_m\cdot C_l^\dagger C_n)
   \nonumber\\
 &=& \tr_B(\hat{M}_i\,\hat{M}_j^\dagger)\,,
\label{eq:T24_me}
\eea
where $\hat{M}_i \equiv C_{p_i}^\dagger\,C_{q_i}$ is a
$d_B\times d_B$ matrix for the $i$th pair in the
group~$g_\omega$.
Since $T_{13}$ is Hermitian,
$\<mn|T_{13}|kl\> = (\<kl|T_{13}|mn\>)^*
= [\tr_A(M_i^\dagger M_j)]^*$.
Combining with Eq.~(\ref{eq:T24_me}) and summing
over the diagonal ensemble as in
Sec.~\ref{app:time_avg}, each summand takes the form
$\tr_B(\hat{M}_i\,\hat{M}_j^\dagger)\cdot
[\tr_A(M_i^\dagger M_j)]^*$.
Relabeling $i\leftrightarrow j$ in the double sum and
using the cyclicity of the trace, this can be
rewritten as
\be
\label{eq:I1_grouped}
\avg{I_1^{(2)}} = \sum_\omega
 \sum_{i,j=1}^{N_\omega}
 \tr_A(M_i^\dagger M_j)\cdot
 \tr_B(\hat{M}_i^\dagger \hat{M}_j)\,.
\ee
Unlike $\avg{I_0^{(2)}}$ in
Eq.~(\ref{eq:I0_grouped}), the two trace factors in
each summand are not complex conjugates of each other,
so the expression does not reduce to a modulus squared.
The sum is nonetheless real, since the
$i\leftrightarrow j$ relabeling complex-conjugates
each summand.
Equation~(\ref{eq:I1_grouped}) is equivalent to the
identity~(\ref{eq:I1_I0SU}) after the same
$i\leftrightarrow j$ relabeling.

\subsection{Assembling the time-averaged
entangling power}
\label{app:assemble}

Combining the results above with
Eq.~(\ref{eq:ep_def}), the time-averaged entanglement
power is
\bea
\avg{\ep} &=& 1 - C_{d_A}\,C_{d_B}\,
\Big[\tr(T_{13}) + \avg{I_0^{(2)}}
\nonumber\\
&& \qquad\qquad
 + \;\tr(T_{24}) + \avg{I_1^{(2)}}\Big]\,,
\eea
where $\avg{I_0^{(2)}}$ and $\avg{I_1^{(2)}}$ are
given by Eqs.~(\ref{eq:I0_grouped})
and~(\ref{eq:I1_grouped}), and the constant terms are
$\tr(T_{13}) = d_A\,d_B^2$ and
$\tr(T_{24}) = d_A^2\,d_B$.
This completes the derivation of the algorithm.

\section{Spectral
decomposition of the spin-1 XYZ Hamiltonian}
\label{app:Z2Z2}

In this appendix we derive the symmetry decomposition
used in Sec.~\ref{sec:isospectral} to establish the
uniqueness of the XXX degeneracy pattern for
spin-1.

\subsection{Discrete symmetry}

The general XYZ Hamiltonian~(\ref{eq:H_2site}) does not
commute with $S_z^{\rm tot}$ for $a_x \neq a_y$, but
it commutes with the parity operators
$R_\alpha \equiv e^{i\pi S_{\alpha,1}} \otimes
e^{i\pi S_{\alpha,2}}$ for $\alpha = x,y,z$.
A $\pi$-rotation about the $\alpha$-axis preserves
$S_\alpha$ and flips the other two components,
\be
\label{eq:Rconj_app}
e^{-i\pi S_\alpha}\, S_\beta\, e^{i\pi S_\alpha}
= \begin{cases} +S_\beta & \beta = \alpha\,,\\
-S_\beta & \beta \neq \alpha\,,
\end{cases}
\ee
so each bilinear term $S_\beta^{(1)} S_\beta^{(2)}$
picks up two sign flips that cancel, giving
$[R_\alpha, H] = 0$.
The three operators satisfy $R_x R_y R_z = \Id$ and
$R_\alpha^2 = \Id$, generating a
$\mathbb{Z}_2 \times \mathbb{Z}_2$ (Klein four-)
group.

\subsection{Sector decomposition}

For spin-1, $R_z$ has eigenvalue $(-1)^{m_1+m_2}$
on $|m_1, m_2\>$, splitting the nine product states
into a 5-dimensional even and a 4-dimensional odd
sector.
The operator $R_x$ acts as
$|m_1,m_2\> \to |-m_1,-m_2\>$ and commutes with $R_z$.
Forming symmetric and antisymmetric combinations
under $R_x$ within each $R_z$ sector yields four blocks.
Writing $S_x$ and $S_y$ in terms of
$S_\pm = S_x \pm iS_y$ and using
$H = \frac{a_x+a_y}{4}(S_+^{(1)}S_-^{(2)}
+ S_-^{(1)}S_+^{(2)})
+ \frac{a_x-a_y}{4}(S_+^{(1)}S_+^{(2)}
+ S_-^{(1)}S_-^{(2)})
+ a_z\, S_z^{(1)}S_z^{(2)}$,
we compute the Hamiltonian in each sector.

\paragraph{$(R_z,R_x) = (+,+)$, dimension 3.}
In the basis
$\{|0,0\>,\;(|1,1\>+|-\!1,-\!1\>)/\!\sqrt{2},\;
(|1,-\!1\>+|-\!1,1\>)/\!\sqrt{2}\}$,
\be
\label{eq:Hpp_app}
H_{(+,+)} \!=\!
\begin{pmatrix}
0 & \frac{a_x-a_y}{\sqrt{2}} & \frac{a_x+a_y}{\sqrt{2}} \\[4pt]
\frac{a_x-a_y}{\sqrt{2}} & a_z & 0 \\[4pt]
\frac{a_x+a_y}{\sqrt{2}} & 0 & -a_z
\end{pmatrix}\!.
\ee
This matrix is traceless, so its characteristic
polynomial is the depressed
cubic polynomial in Eq.~(\ref{eq:depressed_cubic}).

\paragraph{$(R_z,R_x) = (+,-)$, dimension 2.}
In the basis
$\{(|1,1\>-|-\!1,-\!1\>)/\!\sqrt{2},\;
(|1,-\!1\>-|-\!1,1\>)/\!\sqrt{2}\}$,
\be
\label{eq:Hpm_app}
H_{(+,-)} =
\begin{pmatrix} a_z & 0 \\ 0 & -a_z \end{pmatrix}\,,
\ee
since the only states that could mediate off-diagonal
coupling ($|0,0\>$) belong to the $(+,+)$ sector.
The eigenvalues are $\pm a_z$.

\paragraph{$(R_z,R_x) = (-,+)$, dimension 2.}
In the basis
$\{(|1,0\>+|-\!1,0\>)/\!\sqrt{2},\;
(|0,1\>+|0,-\!1\>)/\!\sqrt{2}\}$,
\be
\label{eq:Hmp_app}
H_{(-,+)} =
\begin{pmatrix} 0 & a_x \\ a_x & 0 \end{pmatrix}\,,
\ee
with eigenvalues $\pm a_x$.  (The diagonal vanishes
because $S_z^{(1)}S_z^{(2)} = 0$ when one $m$-value
is zero, and the off-diagonal element receives equal
contributions
$\frac{a_x+a_y}{2}$ and $\frac{a_x-a_y}{2}$ from the
flip-flop and $\Delta m = \pm 2$ terms, summing to
$a_x$.)

\paragraph{$(R_z,R_x) = (-,-)$, dimension 2.}
In the basis
$\{(|1,0\>-|-\!1,0\>)/\!\sqrt{2},\;
(|0,1\>-|0,-\!1\>)/\!\sqrt{2}\}$,
\be
\label{eq:Hmm_app}
H_{(-,-)} =
\begin{pmatrix} 0 & a_y \\ a_y & 0 \end{pmatrix}\,,
\ee
with eigenvalues $\pm a_y$.  The off-diagonal element
now receives contributions $\frac{a_x+a_y}{2}$ and
$-\frac{a_x-a_y}{2}$ (the sign flip from the
antisymmetric combination), summing to $a_y$.

\vspace{6pt}
\subsection{Uniqueness of the $\{1,3,5\}$ pattern}

The full spectrum consists of the three roots of
$p(\lambda)$ together with $\pm a_x$, $\pm a_y$,
$\pm a_z$.  For a 5-fold degenerate eigenvalue
$\lambda_0$ to exist, it must appear in all four sectors.
From the two-dimensional sectors,
$\lambda_0 \in \{\pm a_x\} \cap \{\pm a_y\}
\cap \{\pm a_z\}$, which requires
$|a_x| = |a_y| = |a_z|$.
The identity
$p(a_x) = -a_x(a_y - a_z)^2$
shows $a_x$ is a root of $p$ if and only if $a_y = a_z$,
with analogous identities for cyclic permutations.
Combining these constraints forces
$a_x = a_y = a_z$---the XXX point
(up to sign flips from local unitaries).

\section{Analytic entangling power for the spin-1 XXZ
model}
\label{app:xxz_analytic}

In this appendix we derive closed-form expressions
for both the instantaneous entangling power $\ep(t)$
and its time average $\avg{\ep}$ for the
two-site spin-1 XXZ Hamiltonian,
\be
\label{eq:H_XXZ_app}
H = S_x^{(1)}S_x^{(2)} + S_y^{(1)}S_y^{(2)}
    + \Delta\, S_z^{(1)}S_z^{(2)}\,,
\ee
valid for generic values of $\Delta$.
The calculation uses the algorithm of
Appendix~\ref{app:algorithm}, together with the
$\mathbb{Z}_2\times\mathbb{Z}_2$ decomposition of
Appendix~\ref{app:Z2Z2}.
We first obtain $\ep(t)$ as an explicit trigonometric
polynomial, then recover $\avg{\ep}$ by
time averaging.

\subsection{Spectrum and eigenstates}

Setting $a_x = a_y = 1$ and $a_z = \Delta$ in the
sector Hamiltonians of Appendix~\ref{app:Z2Z2},
the $(+,-)$ block~(\ref{eq:Hpm_app}) gives eigenvalues
$\pm\Delta$, the $(-,+)$ and $(-,-)$
blocks~(\ref{eq:Hmp_app})--(\ref{eq:Hmm_app}) each
give $\pm 1$ (since $a_x = a_y = 1$), and the
$3\times 3$ $(+,+)$ block~(\ref{eq:Hpp_app})
reduces to
\be
\label{eq:Hpp_xxz}
H_{(+,+)} = \begin{pmatrix}
0 & 0 & \sqrt{2} \\
0 & \Delta & 0 \\
\sqrt{2} & 0 & -\Delta
\end{pmatrix}\,,
\ee
whose characteristic polynomial is
$\lambda^3 - (\Delta^2 + 8)\lambda/4 = 0$,
giving roots $0$ and
$E_\pm \equiv (-\Delta \pm \varsigma)/2$
where $\varsigma \equiv \sqrt{\Delta^2 + 8}$.
The full spectrum is
\be
\label{eq:xxz_spectrum}
\{E_n\} = \{\Delta,\,\Delta,\,
1,\,-1,\,1,\,-1,\,-\Delta,\,E_+,\,E_-\}\,,
\ee
listed in the sector order of
Appendix~\ref{app:Z2Z2}.
The corresponding coefficient matrices
$C_n$ ($3\times 3$, with rows labeled by the
first-site $m$-value and columns by the second)
are
\bea
\label{eq:Cn_rational}
C_0 &=& |1,1\>\,,\quad
C_1 = |-\!1,-\!1\>\,,
\quad C_6 = \tfrac{|1,-\!1\>-|-\!1,1\>}{\sqrt{2}}\,,
\nonumber\\
C_{2,3} &=& \tfrac{|1,0\>\pm|0,1\>}{\sqrt{2}}\,,
\quad
C_{4,5} = \tfrac{|0,-\!1\>\pm|-\!1,0\>}{\sqrt{2}}\,,
\nonumber
\eea
for the seven ``rational'' states ($E = \Delta$,
$\pm 1$, $-\Delta$), and
\bea
\label{eq:Cn_irrational}
C_{7,8} &=& \frac{1}{N_\pm}\Big(\sqrt{2}\,|0,0\>
\nonumber\\
&&\qquad +\;\frac{E_\pm}{\sqrt{2}}
\big(|1,-\!1\>+|-\!1,1\>\big)\Big)
\eea
for the $E_\pm$ states in the $(+,+)$ sector, with
normalization $N_\pm^2 = 2 + E_\pm^2
= \varsigma(\varsigma \mp \Delta)/2$.
The sparsity of these matrices---each has at most
two nonzero entries---is the key to making the
trace computation tractable.

\subsection{$\omega$-group structure}

For generic $\Delta$
($\Delta \notin \{0,\,\pm 1,\,\pm 2\}$),
all nine eigenvalues are distinct and the $d^4 = 81$
pairs $(k,l)$ partition into 27~$\omega$-groups.
Of these, 15 groups involve only the rational states
(0--6) and have $\Delta$-independent
$M$-matrices.  They decompose as follows:
one $\omega = 0$ group (containing 15 pairs,
since each of the 9 diagonal pairs has $\omega = 0$
and 6 off-diagonal pairs from the three
two-fold degeneracies);
two groups at $\omega = \pm 2$ (4 pairs each,
from $E_k = 1$, $E_l = -1$ and vice versa);
two at $\omega = \pm 2\Delta$ (2 pairs each);
and four at $\omega = \pm(\Delta \pm 1)$
(6 pairs each).
The remaining 12 groups each contain
1 or 2 pairs involving at least one $E_\pm$
state.  These are the ``irrational'' groups
whose traces live in $\mathbb{Q}(\Delta, \varsigma)$.

\subsection{Algebraic structure of the traces}

Each trace $\tr_A(M_i^\dagger M_j)$
involves bilinear products of $C$-matrix entries.
For the rational states, these products are in
$\mathbb{Q}(\sqrt{2})$ and are $\Delta$-independent.
For the $E_\pm$ states, the individual entries
$\alpha_\pm = \sqrt{2}/N_\pm$ and
$\beta_\pm/\sqrt{2} = E_\pm/(\sqrt{2}\,N_\pm)$
involve irrational normalization factors, but
in the traces the square roots cancel and the
bilinear products lie in the quadratic extension
$\mathbb{Q}(\Delta, \varsigma)$ subject to
$\varsigma^2 = \Delta^2 + 8$.  For example,
\be
\alpha_\pm^2 = \frac{4}{\varsigma(\varsigma \mp
\Delta)}\,,\quad
\alpha_+\,\alpha_- = \frac{\sqrt{2}}{\varsigma}\,.
\ee
The computation then proceeds entirely within
$\mathbb{Q}(\Delta,\varsigma)$, with
$\varsigma^2 \to \Delta^2 + 8$ imposed after
each intermediate step.

\subsection{The parity factor}
\label{app:parity_factor}

The instantaneous entangling power
involves the sum $I_0(t) + I_1(t)$, each containing
all $d^4$ quadruples $(k,l,k',l')$ weighted by the
phase $e^{-i\Omega t}$ with
$\Omega = (E_k - E_l) - (E_{k'} - E_{l'})$.
Since every eigenstate has definite site-exchange
parity, $C_n^T = \epsilon_n\, C_n$ with
$\epsilon_n = \pm 1$, the $\hat{M}$-matrix
(defined as $\hat{M}_{kl} = C_k^T C_l$) is related
to $M_{kl} = C_k C_l^T$ by
$\hat{M}_{kl} = \epsilon_k\,\epsilon_l\; M_{kl}$.
Writing $\sigma_{kl} \equiv \epsilon_k\,\epsilon_l$:
\be
\label{eq:trB_sign}
\tr_A(\hat{M}_{kl}^\dagger \hat{M}_{k'l'})
= \sigma_{kl}\,\sigma_{k'l'}\;
\tr_A(M_{kl}^\dagger M_{k'l'})\,.
\ee
The sum $I_0 + I_1$ then acquires a
\emph{parity factor}:
\bea
\label{eq:parity_factor}
&&\hspace{-0.7cm}
I_0(t) + I_1(t)
= 2d^3
\non\\
&&
+\!\sum_{k,l,k',l'}\!
e^{-i\Omega t}\,
(1 + \sigma_{kl}\sigma_{k'l'})\,
[\tr_A(M_{kl}^\dagger M_{k'l'})]^2.
\eea
This factor takes the value~2 when both pairs
share the same exchange parity
($\sigma_{kl} = \sigma_{k'l'}$) and vanishes
otherwise: contributions from parity-mismatched
pairs cancel between $I_0$ and $I_1$.

The exchange parities of the nine eigenstates are
\be
\label{eq:eps_values}
\{\epsilon_n\} = \{+,+,+,-,+,-,-,+,+\}\,,
\ee
where states~0, 1, 2, 4, 7, 8 are symmetric
($C^T = C$) and states~3, 5, 6 are antisymmetric
($C^T = -C$).
One can verify that \emph{within} each
$\omega$-group, all pairs $(k,l)$ share the
same value of $\sigma_{kl}$, so the parity factor
is uniformly~2 for same-group terms.  This is
why $\avg{I_0} = \avg{I_1}$ after time averaging,
which restricts to same-group contributions.
For cross-group terms ($\Omega \neq 0$),
pairs from groups with different $\sigma$ values
are killed by the parity factor.

\subsection{The instantaneous entangling power}
\label{app:ep_instant}

Since all coefficient matrices are real, every trace
is real and $I_0(t) + I_1(t)$ is a real trigonometric
polynomial.  The contributions at $+\Omega$ and
$-\Omega$ combine into cosines, giving
\be
\label{eq:ep_t_app}
\ep(t) = \frac{5}{8}
- \frac{1}{144}\bigg[
\mathcal{A}_0
+ \sum_{j=1}^{24} \mathcal{A}_j\,
\cos(\Omega_j\, t)\bigg]\,,
\ee
with $5/8 = 1 - 2d^3/[d^2(d+1)^2]$ for $d = 3$.
The DC component is
\be
\label{eq:A0_app}
\mathcal{A}_0
= \frac{46\Delta^4 + 684\Delta^2 + 2636}
       {(\Delta^2+8)^2}\,,
\ee
and the 24 oscillating frequencies fall into
8 rational and 16 irrational terms.

\medskip\noindent\textbf{Rational frequencies}
(8 terms).  These arise from beat frequencies
among the rational eigenvalues
$\{\Delta, 1, -1, -\Delta\}$:
\begin{center}
\begin{tabular}{c|c}
$\Omega_j$ & $\mathcal{A}_j$ \\
\hline
$\Delta$ & $4/(\Delta^2+8)$ \\[2pt]
$\Delta + 2$ & $2\Delta^2/(\Delta^2+8)$ \\[2pt]
$\Delta - 2$ & $2\Delta^2/(\Delta^2+8)$ \\[2pt]
$2\Delta + 2$ & $4$ \\[2pt]
$2\Delta - 2$ & $4$ \\[2pt]
$3\Delta$ & $8/(\Delta^2+8)$ \\[2pt]
$4\Delta$ & $2$ \\[2pt]
$4$ & $3$
\end{tabular}
\end{center}

\medskip\noindent\textbf{Irrational frequencies}
(16 terms).  These involve
$\varsigma = \sqrt{\Delta^2 + 8}$ and come in
conjugate pairs under
$\varsigma \to -\varsigma$:
\begin{table}[h]
\centering
\begin{tabular}{c|c}
$\Omega_j$ & $\mathcal{A}_j$ \\
\hline
$\varsigma$ & $(52\Delta^2 + 272)/(\Delta^2+8)^2$ \\[2pt]
$2\varsigma$ & $36/(\Delta^2+8)^2$ \\[2pt]
$-\Delta \pm \varsigma$ & $(\Delta^2 \pm
  \Delta\varsigma + 4)/[2(\Delta^2+8)]$ \\[2pt]
$3\Delta \pm \varsigma$ & $(\Delta^2 \pm
  \Delta\varsigma + 4)/(\Delta^2+8)$ \\[2pt]
$(3\Delta \pm \varsigma)/2$ & $(6\Delta^2 \pm
  2\Delta\varsigma + 48)/(\Delta^2+8)$ \\[3pt]
$\Delta \pm \varsigma + 2$ & $8/(\Delta^2+8)$ \\[2pt]
$\Delta \pm \varsigma - 2$ & $8/(\Delta^2+8)$ \\[2pt]
$(\Delta \pm \varsigma)/2 + 2$ & $(3\Delta^2 \pm
  \Delta\varsigma + 24)/(\Delta^2+8)$ \\[3pt]
$(\Delta \pm \varsigma)/2 - 2$ & $(3\Delta^2 \pm
  \Delta\varsigma + 24)/(\Delta^2+8)$
\end{tabular}
\caption{{\em Fourier amplitudes and irrational
frequencies for the instantaneous entangling power
of the spin-1 XXZ model,
Eq.~(\ref{eq:ep_t_app}).
Each ``$\pm$'' row represents two distinct
frequencies (upper and lower signs).
}}
\label{tab:fourier_xxz}
\end{table}

\noindent
As a consistency check, setting all cosines to~1
(corresponding to $t = 0$, i.e., $U = \mathbf{1}$)
gives $\mathcal{A}_0 + \sum_j \mathcal{A}_j = 90$,
so that $\ep(0) = 5/8 - 90/144 = 0$, as required.

\subsection{Time averaging and $\avg{\ep}$}
\label{app:xxz_time_avg}

Time averaging $\ep(t)$ kills all oscillating
terms, leaving only the DC component:
$\avg{\ep} = 5/8 - \mathcal{A}_0/144$.
Equivalently, this corresponds to restricting
the sum in Eq.~(\ref{eq:parity_factor}) to
same-$\omega$-group contributions
($\Omega = 0$), for which the parity factor
is uniformly~2.  This gives
$\avg{I_0} + \avg{I_1} = 2\avg{I_0}$,
i.e., $\avg{I_0} = \avg{I_1}$
(or equivalently,
$\avg{I_0^{(2)}} = \avg{I_1^{(2)}}$,
see Appendix~\ref{app:I1}).

\medskip\noindent\textbf{Cancellation of $\varsigma$.}
Within each $\omega$-group, the traces involving the
$E_\pm$ eigenstates live in the quadratic extension
$\mathbb{Q}(\Delta,\varsigma)$.  However, the 12
irrational groups come in six conjugate pairs related
by $\varsigma \to -\varsigma$
($E_+ \leftrightarrow E_-$).
For example, the pair at
$\omega = (3\Delta \mp \varsigma)/2$ contributes
\be
\label{eq:conj_pair_example}
\frac{\Delta^2 \mp \Delta\,\varsigma + 4}
     {4(\Delta^2 + 8)}
\ee
to $\avg{I_0^{(2)}}$.  Adding the two conjugates:
\be
\frac{\Delta^2 - \Delta\varsigma + 4}{4(\Delta^2+8)}
+ \frac{\Delta^2 + \Delta\varsigma + 4}{4(\Delta^2+8)}
= \frac{\Delta^2 + 4}{2(\Delta^2+8)}\,,
\ee
a rational function of $\Delta$.
The same cancellation occurs for every conjugate
pair, so all $\varsigma$-dependence drops out of
$\avg{I_0^{(2)}}$.

\medskip\noindent\textbf{Result.}
Summing the 27~$\omega$-group contributions and
adding the static part $\tr(T_{13}) = d^3 = 27$:
\be
\label{eq:I0_xxz}
\avg{I_0^{(2)}} + \tr(T_{13})
= \frac{23\Delta^4 + 342\Delta^2 + 1318}{(\Delta^2+8)^2}
+ 27\,,
\ee
with $\avg{I_1^{(2)}} + \tr(T_{24})$ equal to the
same value.
Inserting into
$\avg{\ep} = 1 - [\avg{I_0^{(2)}} + \tr(T_{13})
+ \avg{I_1^{(2)}} + \tr(T_{24})]\,/\,[d^2(d+1)^2]$
with $d = 3$
gives Eq.~(\ref{eq:ep_xxz}) in the main text.
This expression is manifestly positive, symmetric
under $\Delta \to -\Delta$, and monotonically
increasing toward $11/36$ as $|\Delta| \to \infty$.
It has been verified numerically at
$\Delta = 1/2$ (yielding $4421/13068$) and
$\Delta = 3$ (yielding $3373/10404$), as well
as at $\Delta = 3/2$ and numerous other
non-degenerate values.

It is worth pointing out that the time average
of $\ep(t)$ provides an independent derivation of
$\avg{\ep}$: one simply reads off
the constant term from
Eq.~(\ref{eq:ep_t_app}), bypassing the need to
sum $\omega$-group contributions and cancel
$\varsigma$ explicitly.

\subsection{Validity and degenerate points}

Both Eq.~(\ref{eq:ep_t_app}) and
Eq.~(\ref{eq:ep_xxz}) were derived under
the assumption that the 27 $\omega$-groups are all
distinct.  At special values of $\Delta$ where
eigenvalue differences coincide---namely
$\Delta = 0$, $\pm 1$, and $\pm 2$---some groups merge
and the cross-terms between formerly separate groups
contribute additional positive terms to
$\avg{I_0^{(2)}}$.  Since $\avg{\ep}$ decreases
with increasing $I_0 + I_1$, the true value at
these degenerate points lies \emph{below} the generic
formula.  The magnitude of the drop depends on
the extent of the degeneracy:
at $\Delta = \pm 1$, where the number of groups drops
from 27 to 7, the XXX point has
$\avg{\ep} = 13/54 \approx 0.241$ versus the
generic-formula value of $981/2916 \approx 0.336$;
at $\Delta = \pm 2$, where 27 groups reduce to 25,
the drop is much milder
($\avg{\ep} \approx 0.326$ versus $0.330$).

Equation~(\ref{eq:ep_xxz}) thus provides the
``background curve'' of $\avg{\ep}$ as a smooth
function of the anisotropy.
The sharp dips at
$\Delta = 0$ and $\pm 1$ discussed in the main text
are superimposed on this background as
measure-zero discontinuities of the infinite-time
average, arising from accidental eigenvalue
degeneracies that enlarge the $\omega$-groups.

\section{Bethe Ansatz and the Two-Magnon $S$-matrix}
\label{app:bethe}

This appendix provides a self-contained derivation of
the two-magnon scattering matrix for the spin-$1/2$ XXZ
chain starting from the coordinate Bethe
ansatz~\cite{Bethe:1931hc}.
The presentation follows the standard treatment in
Refs.~\cite{Korepin:1993kvr,Gaudin:2014,Faddeev:1996iy}; we include
it here because the Bethe ansatz framework may be
less familiar to readers approaching spin chains
from the quantum-information perspective.
Following the Bethe ansatz literature, we write
$S_j^\alpha$ for the spin operator at site~$j$
(rather than $S_{\alpha,j}$ used in the main text).

\subsection{Setup and the ferromagnetic vacuum}
\label{app:bethe_setup}

We consider the spin-$1/2$ XXZ Hamiltonian on a
periodic chain of $L$ sites:
\be
H_{\rm XXZ} = \sum_{j=1}^{L}
\bigl(S^x_j S^x_{j+1} + S^y_j S^y_{j+1}
+ \Delta\, S^z_j S^z_{j+1}\bigr)\,,
\label{eq:H_xxz_bethe}
\ee
with periodic boundary conditions
$\vec{S}_{L+1} \equiv \vec{S}_1$ and
anisotropy parameter $\Delta$.
We parametrize $\Delta = \cos\gamma$ with
$\gamma \in [0,\pi]$.

The Hamiltonian conserves the total spin projection
$S^z_{\rm tot} = \sum_j S^z_j$, which allows
diagonalization within sectors of fixed $M$ down-spins.
The fully polarized state
$|\Omega\rangle \equiv |\!\uparrow\uparrow
\cdots\uparrow\rangle$
is an eigenstate with energy
$E_0 = \tfrac{1}{4}L\Delta$,
and serves as the reference vacuum.
Spin-flip excitations above this vacuum are called
magnons.

\subsection{One-magnon sector}
\label{app:one_magnon}

A single spin flip at site $n$ creates the state
$|n\rangle \equiv S^-_n |\Omega\rangle$.
Acting with the Hamiltonian on the ansatz
$|\psi\rangle = \sum_{n=1}^{L} \phi(n)\,|n\rangle$
yields the eigenvalue equation
\begin{multline}
\tfrac{1}{2}[\phi(n+1) + \phi(n-1)]
+ \tfrac{1}{4}(L-2)\Delta\,\phi(n)  \\
- \tfrac{1}{2}\Delta\,\phi(n) = E\,\phi(n)\,.
\label{eq:one_magnon_eig}
\end{multline}
Substituting the plane-wave ansatz
$\phi(n) = e^{ipn}$ immediately gives the
magnon dispersion relation
\be
\epsilon(p) \equiv E - E_0
= \Delta - \cos p\,,
\label{eq:magnon_disp}
\ee
where $\epsilon(p)$ is the magnon energy measured
relative to the vacuum.
The quasi-momentum $p$ is quantized by the
periodic boundary condition:
$e^{ipL} = 1$, so $p = 2\pi m/L$ with
$m = 0, 1, \ldots, L-1$.

\subsection{Two-magnon sector and the Bethe wavefunction}
\label{app:two_magnon}

Two magnons at positions $n_1 < n_2$ define the
state $|n_1, n_2\rangle = S^-_{n_1} S^-_{n_2}
|\Omega\rangle$.
In the region $n_2 - n_1 \geq 2$, where the two
flipped spins do not occupy adjacent sites, each magnon
propagates independently and the eigenvalue equation
reduces to two copies of the one-magnon problem.
The Bethe ansatz wavefunction in this region is a
superposition of incoming and outgoing plane waves:
\be
\psi(n_1, n_2)
= A\, e^{i(p_1 n_1 + p_2 n_2)}
+ B\, e^{i(p_2 n_1 + p_1 n_2)}\,,
\label{eq:bethe_wf}
\ee
where $p_1$ and $p_2$ are the quasi-momenta of the
two magnons, $A$ is the amplitude for the ``direct''
configuration (magnon 1 to the left with momentum $p_1$),
and $B$ is the amplitude for the ``exchanged''
configuration (momenta swapped).
The total energy is the sum of individual magnon
energies:
\be
E - E_0 = \epsilon(p_1) + \epsilon(p_2)
= 2\Delta - \cos p_1 - \cos p_2\,.
\ee

The nontrivial physics arises when the two magnons
are adjacent, $n_2 = n_1 + 1$.
In this case the exchange terms in the Hamiltonian
produce a contact interaction: the eigenvalue equation
at the boundary $n_2 = n_1 + 1$ differs from the bulk
equation by an additional term proportional to $\Delta$.
Substituting the ansatz~(\ref{eq:bethe_wf}) into the
boundary condition and requiring consistency with the
bulk equation yields the scattering amplitude ratio
\be
\frac{B}{A}
= -\frac{e^{i(p_1 + p_2)} - 2\Delta\,e^{ip_2} + 1}
       {e^{i(p_1 + p_2)} - 2\Delta\,e^{ip_1} + 1}
\equiv -e^{i\phi(p_1, p_2)}\,,
\label{eq:scat_phase}
\ee
where $\phi(p_1, p_2)$ is the two-magnon scattering phase.
The key observation of Bethe~\cite{Bethe:1931hc} is that
this ratio is a \emph{pure phase}: $|B/A| = 1$.
Physically, this means magnon scattering in the XXZ chain
is purely elastic---magnons exchange momenta but are
never reflected or absorbed.
This is the hallmark of integrability.

The quantization conditions on a periodic chain of
length $L$ are the \emph{Bethe equations}:
\be
e^{ip_1 L} = -\frac{B}{A}\,,
\qquad
e^{ip_2 L} = -\frac{A}{B}\,,
\label{eq:bethe_eqs}
\ee
which state that the phase accumulated by each magnon
around the chain equals the product of all
pairwise scattering phases.

\subsection{Spectral parameter, rapidity, and the $R$-matrix}
\label{app:rapidity}

The scattering phase~(\ref{eq:scat_phase}) can be
greatly simplified by trading the quasi-momenta $p_j$
for the trigonometric \emph{spectral parameter}~$u_j$,
defined by~\cite{Korepin:1993kvr,Faddeev:1996iy}
\be
e^{ip_j}
= \frac{\sin(u_j + \gamma/2)}{\sin(u_j - \gamma/2)}\,.
\label{eq:spectral_def}
\ee
Substituting into the scattering
amplitude~(\ref{eq:scat_phase}) yields, after
straightforward algebra,
\be
\frac{B}{A}
= \frac{\sin(u_{12} - \gamma)}
       {\sin(u_{12} + \gamma)}\,,
\label{eq:BA_spectral}
\ee
where $u_{12} \equiv u_1 - u_2$.
This depends only on the \emph{difference} of the
spectral parameters, which is the hallmark of
integrability: the scattering is a function of
the relative rapidity alone.

In the isotropic limit $\gamma \to 0$
($\Delta \to 1$), the spectral parameter reduces
to the rational Bethe rapidity.
Writing $u_j = \gamma\lambda_j$ and taking
$\gamma \to 0$ in Eq.~(\ref{eq:spectral_def})
gives
$e^{ip_j} \to (\lambda_j + \tfrac{1}{2}i)/
(\lambda_j - \tfrac{1}{2}i)$,
which inverts to
$\lambda_j = \tfrac{1}{2}\cot(p_j/2)$.
In this limit Eq.~(\ref{eq:BA_spectral}) becomes
$B/A \to (\lambda_{12} - i)/(\lambda_{12} + i)$,
recovering the familiar rational form of the
XXX scattering phase.

To construct the full two-body scattering matrix,
we note that the XXZ Hamiltonian conserves
$S^z_{\rm tot}$.
In the two-magnon sector, the Hilbert space at
each pair of sites decomposes into sectors of
total $S^z = +1, 0, -1$.
The $S^z = \pm 1$ sectors (both spins aligned)
are one-dimensional and trivially scatter with
unit amplitude.
The $S^z = 0$ sector is two-dimensional, spanned
by $|\!\uparrow\downarrow\rangle$ and
$|\!\downarrow\uparrow\rangle$, and it is here
that the nontrivial scattering takes place.

Writing the $S$-matrix on $\CC^2 \otimes \CC^2$
in the computational basis
$\{|\!\uparrow\uparrow\rangle,
|\!\uparrow\downarrow\rangle,
|\!\downarrow\uparrow\rangle,
|\!\downarrow\downarrow\rangle\}$
yields the six-vertex $R$-matrix:
\be
R(u, \gamma)
= \begin{pmatrix}
a & 0 & 0 & 0 \\
0 & b & c & 0 \\
0 & c & b & 0 \\
0 & 0 & 0 & a
\end{pmatrix}\,,
\label{eq:R_6v_app}
\ee
with vertex weights
$a = \sin(u + \gamma)$, $b = \sin u$,
$c = \sin\gamma$,
where $u$ is the spectral-parameter difference
$u_{12}$ defined above.
The name ``six-vertex'' refers to the six nonzero
matrix elements, which correspond to the six
allowed arrow configurations in the ice-type
vertex model~\cite{Lieb:1967zz,Lieb:1967b,Baxter:1972hz}.
The anisotropy is encoded by
$\Delta = \cos\gamma$.

\subsection{Yang--Baxter equation}
\label{app:ybe}

The integrability of the XXZ chain is ultimately
guaranteed by the Yang--Baxter
equation~\cite{Yang:1967bm,Baxter:1972hz,Sklyanin:1979}:
\begin{multline}
R_{12}(u_{12})\, R_{13}(u_{13})\, R_{23}(u_{23})  \\
= R_{23}(u_{23})\, R_{13}(u_{13})\, R_{12}(u_{12})\,,
\label{eq:ybe}
\end{multline}
where $u_{ij} \equiv u_i - u_j$ and $R_{jk}$ acts
on spaces $j$ and $k$.
Equation~(\ref{eq:ybe}) is a matrix equation in
$(\CC^2)^{\otimes 3}$ and states that the order
of pairwise scatterings in a three-body collision
does not affect the outcome.

The physical consequence is
factorized scattering~\cite{Zamolodchikov:1978xm,Faddeev:1996iy}:
the $N$-magnon $S$-matrix decomposes into a product
of $\binom{N}{2}$ two-body $S$-matrices, and
the Bethe equations~(\ref{eq:bethe_eqs}) generalize
to $N$ coupled equations whose consistency is
precisely the Yang--Baxter equation.
This is the algebraic foundation of integrability
in the XXZ chain.

We note in passing that the quantum group symmetry
$U_q(\mathfrak{sl}_2)$ discussed in
Sec.~\ref{sec:root_of_unity}
is intimately connected to the $R$-matrix: the
Yang--Baxter equation is the defining relation
of the quasi-triangular Hopf algebra structure
of $U_q(\mathfrak{sl}_2)$, and the six-vertex
$R$-matrix is its fundamental
representation~\cite{Drinfeld:1985,Jimbo:1985zk,Chari:1994pz}.

\subsection{From $R$-matrix to physical $S$-matrix}
\label{app:unitarize}

The $R$-matrix~(\ref{eq:R_6v_app}) is not unitary
for real spectral parameter $u$, as the vertex
weights are in general complex.
To obtain the physical unitary $S$-matrix appropriate
for real-time scattering, we analytically continue
$u \to i\theta$, where $\theta$ is the rapidity
difference of the two magnons.
This yields the vertex weights
\begin{align}
a &= \sin(i\theta + \gamma) = i\sinh\theta\,\cos\gamma
+ \cosh\theta\,\sin\gamma\,, \notag \\
b &= \sin(i\theta) = i\sinh\theta\,, \notag \\
c &= \sin\gamma\,.
\end{align}
Normalizing by
$\rho \equiv \sqrt{\sin^2\!\gamma + \sinh^2\!\theta}$
yields the unitary $S$-matrix
$S(\theta, \gamma) = R(i\theta, \gamma)/\rho$,
which is the matrix studied in
Sec.~\ref{sec:twomagnon}.
The three eigenvalue phases $\varphi_+$,
$\varphi_t$, and $\varphi_s$ discussed in
Sec.~\ref{sec:ep_smatrix} are the arguments of the
eigenvalues of this unitary matrix.

\subsection{Entangling power of the $S$-matrix}
\label{app:ep_smatrix_derivation}

We derive the entangling power formula
Eq.~(\ref{eq:ep_main}) for any $U(1)$-conserving
two-qubit unitary with eigenvalue phase differences
$\delta_{t+}$ and $\delta_{s+}$.
For $d_A = d_B = 2$ the identity
$1 - \tr(\rho_A^2) = 2\det\rho_A$ simplifies the
entangling power to
\be
\label{eq:ep_det}
\ep(U) = 2\!\int\!\dd\mu(\psi)\,\dd\mu(\phi)
\;\det\rho_A\,,
\ee
where $\dd\mu(\psi)$ and $\dd\mu(\phi)$ denote
independent Haar measures on $\CC^2$.

Since the overall phase $\varphi_+$ drops out of
the entangling power, the $S$-matrix in the
computational basis takes the effective form
$\text{diag}(1,\,B,\,B,\,1)$ with off-diagonal
entries in the $S^z = 0$ block given by the
spin-exchange amplitude $C$, where
$B = (e^{i\delta_{t+}} + e^{i\delta_{s+}})/2$,
$C = (e^{i\delta_{t+}} - e^{i\delta_{s+}})/2$, and
$|B|^2 + |C|^2 = 1$, $\text{Re}(B^*C) = 0$.
Parametrize the Haar-random input states on the
Bloch sphere as
$|\psi\rangle = \cos\tfrac{\alpha}{2}\,|0\rangle
+ e^{i\xi}\sin\tfrac{\alpha}{2}\,|1\rangle$ and
$|\phi\rangle = \cos\tfrac{\beta}{2}\,|0\rangle
+ \sin\tfrac{\beta}{2}\,|1\rangle$,
with polar angles $\alpha, \beta \in [0,\pi]$ and
azimuthal angle $\xi \in [0,2\pi)$,
where the $U(1)$ symmetry has been exploited to fix
the azimuthal angle of $|\phi\rangle$.
Writing $p = \cos^2\!\tfrac{\alpha}{2}$ and
$q = \cos^2\!\tfrac{\beta}{2}$ for compactness,
the output state vector $U|\psi\rangle|\phi\rangle$
has components whose reduced density matrix
$\rho_A$ satisfies
\be
\label{eq:detM}
\det\rho_A = \bigl|e^{i\xi}\,w\,F
+ G\bigl[p(1-q) + e^{2i\xi}(1-p)q\bigr]
\bigr|^2,
\ee
where $w = \sqrt{p(1-p)\,q(1-q)}
= \tfrac{1}{4}\sin\alpha\sin\beta$ and
\bea
F &=& 1 - e^{i(\delta_{t+}+\delta_{s+})}\cos(\delta_{t+}-\delta_{s+})\,,
\label{eq:F_def} \\
G &=& -\tfrac{i}{2}\,e^{i(\delta_{t+}+\delta_{s+})}\sin(\delta_{t+}-\delta_{s+})\,.
\label{eq:G_def}
\eea

The Haar measure on each Bloch sphere is
$\dd\mu = \frac{1}{4\pi}\sin\alpha\,\dd\alpha\,
\dd\xi$, so the integral~(\ref{eq:ep_det}) becomes
$\ep = 2\int_0^\pi\!\frac{\sin\alpha\,\dd\alpha}{2}
\int_0^\pi\!\frac{\sin\beta\,\dd\beta}{2}
\int_0^{2\pi}\!\frac{\dd\xi}{2\pi}\,
\det\rho_A$,
i.e.\ a uniform average over $p$, $q$, and $\xi$.
A key simplification is that all cross-terms
between the $F$ and $G$ contributions contain
factors of $e^{\pm i\xi}$ or $e^{\pm 3i\xi}$,
which vanish upon $\xi$-integration.
This leaves
\be
\label{eq:ep_FG}
\ep = 2\left[\frac{|F|^2}{36}
+ \frac{2|G|^2}{9}\right],
\ee
where we have used the Bloch-sphere moments
$\<p(1-p)\> = 1/6$ and $\<p^2\> = 1/3$,
with $\<\cdot\>$ denoting the average over the
uniform measure on $[0,1]$.

Substituting
$|F|^2 = 1 - 2\cos(\delta_{t+}\!+\!\delta_{s+})\cos(\delta_{t+}\!-\!\delta_{s+})
+ \cos^2\!(\delta_{t+}\!-\!\delta_{s+})$
and $|G|^2 = \sin^2\!(\delta_{t+}\!-\!\delta_{s+})/4$, we obtain
\bea
\ep(\delta_{t+},\delta_{s+}) &=& \frac{1}{18}\Bigl[
1 - 2\cos(\delta_{t+}\!+\!\delta_{s+})\cos(\delta_{t+}\!-\!\delta_{s+})
\non\\
&&\quad + \cos^2\!(\delta_{t+}\!-\!\delta_{s+})
+ 2\sin^2\!(\delta_{t+}\!-\!\delta_{s+})\Bigr]
\non\\
&=& \frac{1}{18}\bigl[3
- 2\cos(\delta_{t+}\!+\!\delta_{s+})\cos(\delta_{t+}\!-\!\delta_{s+})
\non\\
&&\quad - \cos^2\!(\delta_{t+}\!-\!\delta_{s+})\bigr]\,,
\label{eq:ep_derivation_result}
\eea
which is Eq.~(\ref{eq:ep_main}).

\bibliographystyle{apsrev4-2}
\bibliography{ep_spin_chain}

\end{document}